\documentclass[journal,twocolumn,10pt,twoside]{IEEEtran}
\usepackage{amsfonts}
\usepackage{amsmath,cases,amssymb}
\usepackage[english]{babel}
\usepackage{amsthm} 
\usepackage[dvips]{graphicx}
\usepackage{cite,color}
\usepackage{color}
\usepackage{caption}
\usepackage{subcaption}
\usepackage{epsfig}
\usepackage{epstopdf}
\usepackage{nomencl}
\usepackage{url}
\usepackage{bm}
\usepackage{multirow}
\usepackage{algorithm}
\usepackage{algorithmicx, algpseudocode}

\usepackage{color}
\usepackage{multirow}
\usepackage{mathtools}
\usepackage{etoolbox}

\newtheorem{mypro}{Proposition}

\newtheorem{mydef}{Definition}

\newcommand{\la}{\langle}
\newcommand{\ra}{\rangle}

\begin{document}
\title{A Distributionally Robust Model Predictive Control for Static and Dynamic Uncertainties in Smart Grids}
\author{Qi Li \textit{Student Member, IEEE}, Ye Shi$^{*}$, \textit{Member, IEEE}, Yuning Jiang, \textit{Member, IEEE}, Yuanming Shi, \textit{Senior Member, IEEE}, Haoyu Wang, \textit{Senior Member, IEEE}, and H. Vincent Poor, \textit{Life Fellow, IEEE} 
\thanks{This work was supported by National Natural Science Foundation of China under Grant 62303319, Shanghai Sailing Program under Grant 22YF1428800. This work was also supported by Grants from the C3.a1 Digital Transformation Institute and the Princeton School of Engineering and Applied Science. (Corresponding authors: Ye Shi)} 
\thanks{Qi Li is with the Institute of Mathematical Sciences, ShanghaiTech University, and Ye Shi, Yuanming Shi, Haoyu Wang are with the School of Information Science and Technology, ShanghaiTech University. (email:{\tt liqi2, shiye, shiym, wanghy@shanghaitech.edu.cn})}
\thanks{Yuning Jiang is with the Automatic Control Laboratory, EPFL, Switzerland. (email: {\tt yuning.jiang@ieee.org})} 
\thanks{H. Vincent Poor is with the Department of Electrical Engineering, Princeton
University, Princeton, NJ 08544 USA (\tt poor@princeton.edu).}
}
\maketitle
\setlength\abovedisplayskip{2.5pt}
\setlength\belowdisplayskip{2.5pt}

\begin{abstract}
The integration of various power sources, including renewables and electric vehicles, into smart grids is expanding, introducing uncertainties that can result in issues like voltage imbalances, load fluctuations, and power losses. These challenges negatively impact the reliability and stability of online scheduling in smart grids. Existing research often addresses uncertainties affecting current states but overlooks those that impact future states, such as the unpredictable charging patterns of electric vehicles. To distinguish between these, we term them static uncertainties and dynamic uncertainties, respectively. This paper introduces WDR-MPC, a novel approach that stands for two-stage Wasserstein-based Distributionally Robust (WDR) optimization within a Model Predictive Control (MPC) framework, aimed at effectively managing both types of uncertainties in smart grids. The dynamic uncertainties are first reformulated into ambiguity tubes and then the distributionally robust bounds of both dynamic and static uncertainties can be established using WDR optimization. By employing ambiguity tubes and WDR optimization, the stochastic MPC system is converted into a nominal one. Moreover, we develop a convex reformulation method to speed up WDR computation during the two-stage optimization. The distinctive contribution of this paper lies in its holistic approach to both static and dynamic uncertainties in smart grids. Comprehensive experiment results on IEEE 38-bus and 94-bus systems reveal the method's superior performance and the potential to enhance grid stability and reliability.

\end{abstract}

\makenomenclature
\renewcommand{\nomname}{NOMENCLATURE}

\renewcommand\nomgroup[1]{%
  \item[\bfseries
  \ifstrequal{#1}{A}{Indices and Sets}{%
  \ifstrequal{#1}{C}{Constant}{%
  \ifstrequal{#1}{D}{Decision Variables}{
  \ifstrequal{#1}{R}{Random Variables}{
  }}}}%
]}
\renewcommand{\nompreamble}{Our main notation is defined here for quick reference. Other notation is defined at its first appearance. the inner product $\mathbf{a}^\top\mathbf{b}$ is denoted as $\langle \mathbf{a},\mathbf{b} \rangle$. $||\cdot||$ is the norm operator. Given two sets $\mathbb{X}$ and $\mathbb{S}$, such that $x\in \mathbb{X}$ and $s\in \mathbb{S}$, the Minkowski sum is defined by $\mathbb{X} \oplus \mathbb{S} \triangleq \{x+s: x\in \mathbb{X}, s\in \mathbb{S}\}$, the Pontryagin set difference is defined by $\mathbb{X} \ominus \mathbb{S} \triangleq \{x: x\oplus \mathbb{S} \subseteq \mathbb{X}\}$, and the Cartesian product is defined by $\mathbb{X} \otimes \mathbb{S} \triangleq \{(x,y): x \in \mathbb{X}, y \in \mathbb{S}\}$. $\hat{x}$ and $\Tilde{x}$ respectively denote the sample point and the actual output of forecasting errors. Meanwhile, $\overline{x}$ and $\underline{x}$  represent the upper and lower bounds of $x$, respectively. $x^*$ denotes the conjugate of $x$. The bold $\bm{x}$ signifies the optimization variable, while symbols lacking any modifications stand for constants.}
\nomenclature[A,01]{\(\mathcal{N}\)}{Index sets of buses in the power grid}
\nomenclature[A,02]{\(\mathcal{G}\)}{Index sets of buses connected with normal generators}
\nomenclature[A,03]{\(\mathcal{B}\)}{Index sets of buses connected with charging stations}
\nomenclature[A,04]{\(\mathcal{R}\)}{Index sets of buses connected with renewable energy sources}
\nomenclature[A,05]{\(\mathcal{C}_k\)}{Index sets of children nodes of node $k$}
\nomenclature[A,06]{\(\mathcal{A}_k\)}{Index sets of parent nodes of node $k$}
\nomenclature[A,07]{\(\mathcal{T}\)}{Index sets of time slots in MPC range}

\nomenclature[C,01]{\(P_{k}^{R}\)}{Forecast power output of RESs at node $k$}
\nomenclature[C,02]{\(V_0\)}{Basic voltage of the distribution network}
\nomenclature[C,03]{\(B\)}{Path Indicator Matrix}
\nomenclature[C,04]{\(r_k,x_k\)}{Real and imaginary part of resistance of line $k$}
\nomenclature[C,05]{\(R\)}{Diagonal matrix consisting of line resistances}
\nomenclature[C,06]{\(P_{k}^{Cpred}\)}{Charging demand prediction at charging station $k$}
\nomenclature[C,07]{\(P_{k}^{L}\)}{Active loads at node $k$ }
\nomenclature[C,08]{\(Q_{k}^{L}\)}{Reactive loads at node $k$}
\nomenclature[C,09]{\(\beta\)}{Confidence level}

\nomenclature[D,01]{\(\bm{P}_k^G\)}{Reactive power generation at node $k$}
\nomenclature[D,02]{\(\bm{Q}_k^G\)}{Reactive power generation at node $k$}
\nomenclature[D,02]{\(\bm{\alpha}\)}{Distribution factor of forecast error}
\nomenclature[D,03]{\(\bm{V_k}\)}{Nodal Voltage at node $k$}
\nomenclature[D,04]{\(\bm{P}_{k}^{B}\)}{Charging/discharging power}
\nomenclature[D,05]{\(\bm{SoC}_{k}\)}{State of charge of the battery at node $k$}
\nomenclature[D,06]{\(\bm{P}_{k}\)}{Power flow on the line $k$}

\nomenclature[R,01]{\(\bm{\xi}_k\)}{Forecast error of RESs at node $k$}
\nomenclature[R,02]{\(\bm{\Xi}\)}{Domain of random variables $\bm{\xi}$}
\nomenclature[R,03]{\(\bm{w}\)}{Random disturbance in MPC}
\nomenclature[R,04]{\(e\)}{Cumulative error caused by disturbance in MPC}
\nomenclature[R,05]{\(\bm{\mathcal{W}}\)}{Domain of Random disturbance in MPC}
\nomenclature[R,06]{\(\mathbb{P}\)}{A probability measure/distribution}
\nomenclature[R,07]{\(\mathcal{M}(\Xi)\)}{Sets contains all distributions which is supported at $\Xi$}
\nomenclature[R,08]{\(\mathbb{E}^{\mathbb{P}}\)}{Expectation respect to probability measure $\mathbb{P}$}
\nomenclature[R,09]{\(\mathcal{B}_r\)}{Ambiguity set of probability measure $\mathbb{P}$}
\printnomenclature
\renewcommand\nomgroup[1]{%
  \item[\bfseries
  \ifstrequal{#1}{A}{Indices and Sets}{%
  \ifstrequal{#1}{C}{Constant}{%
  \ifstrequal{#1}{D}{Decision Variables}{
  \ifstrequal{#1}{R}{Random Variables}
  }}}%
]}

\section{Introduction}
The incorporation of various electric power infrastructures, including renewable energy sources (RESs), plug-in electric vehicles (PEVs), and charging stations, has witnessed significant expansion, presenting new challenges for the online scheduling of smart grids \cite{6099519}. These challenges arise from the inherent uncertainties linked to these power facilities, involving forecasting errors of RESs, random charging patterns of PEVs, flexible loads, and fluctuating electricity prices, among other factors.

To address uncertainties like forecasting errors of RESs, various stochastic methods have been developed. Some studies have assumed 
that forecasting errors follow some prescribed distribution \cite{bienstock2014chance}\cite{lubin2015robust}. Nevertheless, these approaches may be inappropriate since the underlying distribution is hard to obtain beforehand. Therefore, recent research works \cite{summers2015stochastic}\cite{xie2017distributionally} have focused on a distributionally robust optimization method called moment-based distributionally robust optimization.  
These methods utilize past RES data to predict the parameters of the distribution but typically do not consider sampling errors from historical data, potentially leading to poor out-of-sample performance. Guo et al. \cite{guo2018data1}\cite{guo2018data2} first introduced Wassertein metric-based distributionally robust (WDR) optimization to tackle RES uncertainties in both distribution networks and transmission networks, demonstrating better out-of-sample performance. Additional research \cite{zhai2022distributionally} extended WDR optimization to distributed scenarios based on the well-known alternating direction method of multipliers (ADMM) \cite{boyd2011distributed}. However, a common issue in the aforementioned WDR optimization is the increasing computational burden with the number of constraints. To alleviate this issue, Duan et al.  \cite{duan2018distributionally} proposed an upper approximation method to accelerate the WDR optimization problem for RES uncertainties in smart grids. 

In parallel, as PEVs become more prevalent in smart grids, researchers \cite{V2G1, V2G2,zeynali2020multi,wang2019joint,wang2020wasserstein} have applied similar WDR optimization methods to tackle the random charging behaviors of PEVs. However, these approaches primarily consider forecasting errors in charging demands and neglect the dynamics of batteries. Consequently, the accumulation of forecasting errors through the battery dynamics may compromise the stability and reliability of real-time smart grids. To mitigate cumulative errors, model predictive control (MPC) frameworks have been introduced to handle uncertainties in PEV charging behaviors or demand response \cite{Shi1, Shi2, Shi3, li2022distributionally, shi2021distributed}. Yet, while MPC offers adaptability to random charging behaviors, it may not fully meet the requirements of system reliability. In fact, uncertainties in smart grids can affect not only current states but also future states through system dynamics like the battery charging process. We categorize uncertainties that impact current states as static uncertainties, while uncertainties that influence future states through dynamic systems are denoted as dynamic uncertainties. Existing research has mainly focused on static uncertainties \cite{li2022distributionally,nguyen2023distributionally}. Our previous work \cite{li2022distributionally} developed a WDR optimization method to handle static uncertainties of renewable energy and PEVs under the framework of normal MPC, but still ignoring the cumulative errors from PEVs' dynamic uncertainties. A recent work \cite{nguyen2023distributionally} provides an MPC method with a distributionally robust optimization at each time slot to handle the random electricity prices and PEVs' charging /discharging behaviors. However, this method only considers static uncertainties and lacks the treatment of dynamic uncertainties and power flow constraints in smart grids. Thus, the handling of dynamic uncertainties in smart grids remains relatively unexplored. 

Handling dynamic uncertainties is significant because random disturbances (dynamic uncertainties) may accumulate through the dynamic system and cause imprecise prediction for future states and improper control actions. Thus, it may raise potential reliability issues in smart grids. 
An intuitive approach to tackle both static and dynamic uncertainties is to implement a stochastic MPC framework. 
Traditional data-driven approaches, such as the Sample Average Approximation (SAA) method \cite{mohajerin2018data}, may exhibit poor out-of-sample performance. Recently, researchers proposed a novel data-driven Distributionally Robust Model Predictive Control for discrete-time linear systems with additive noise \cite{fochesato2022data, hakobyan2022wasserstein}. This approach handles uncertainties in dynamic systems by constructing ambiguity tubes and deriving distributionally robust bounds by a nonconvex optimization to tighten the constraints in the nominal system. Note that this method was primarily developed for simple control problems and has not been studied for comprehensive problems, such as the operation and control in smart grids. Furthermore, the nonconvex nature of the optimization problem and the constraint size growing with the sample size can lead to significant computational burdens. 

\begin{figure}[H]
    \centering
    \includegraphics[width = \columnwidth]{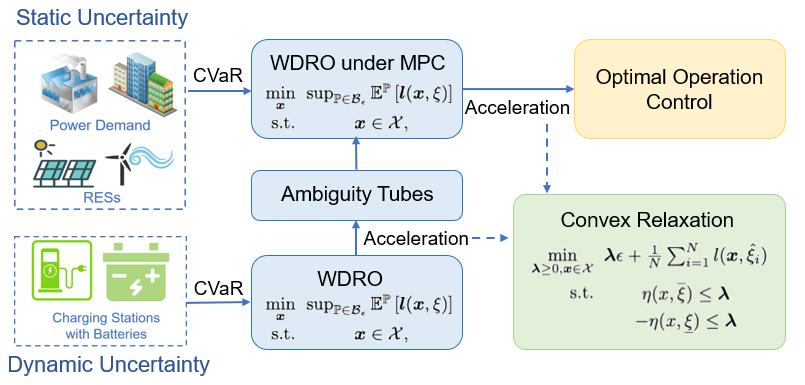}
    \caption{The framework of WDR-MPC in smart grid}
    \label{graph abstract}
\end{figure}

In this paper, we propose a unified Wasserstein-based Distributionally Robust MPC (WDR-MPC) for tackling both static uncertainties and dynamic uncertainties in smart grids. Our method considers the static uncertainties of RESs' forecasting errors and dynamic uncertainties of PEVs' random charging behaviors in a distribution network with comprehensive operational constraints. To address the challenges posed by the two types of uncertainties, we proposed a two-stage algorithm. The first stage involves constructing ambiguity tubes and deriving distributionally robust bounds for dynamic uncertainties using the WDR optimization method, which includes risk function formulations by Conditional Value-at-Risk (CVaR) \cite{rockafellar2000optimization} and convex reformulations. In the second stage, we derive a nominal MPC system from a stochastic MPC system, leveraging the ambiguity tubes for dynamic uncertainties and WDR optimization reformulations for static uncertainties. To alleviate the computational burden associated with WDR optimization, we propose a scalable acceleration method that ensures the constraints do not grow with the sample size. The overall approach is summarized in Figure \ref{graph abstract}. 

Our main contributions are listed as follows, 
\begin{itemize} 
    \item We investigate an online scheduling problem with uncertainties in a distribution network with RESs and charging stations, being the first work to consider both dynamic uncertainties (e.g., random charging behaviors of PEVs) and static uncertainties (e.g., forecasting errors of RESs) in smart grid operation and control. 
    \item We propose WDR-MPC, a novel method that effectively addresses both dynamic and static uncertainties within a unified Wassertein-based optimization framework. Notably, standard WDR optimization methods are inadequate for handling dynamic uncertainties, but our WDR-MPC successfully overcomes this challenge through a two-stage approach. The efficiency of the proposed method is validated by out-of-sample performance tests in real distribution networks.
    \item We provide a tractable convex reformulation for the WDR optimization of two stages and its related acceleration method that is scalable to the sample size. Numerical experiments on real distribution networks demonstrated the scalability of the proposed method.
\end{itemize}

The rest of this paper is organized as follows. Section II introduces preliminaries of the Wasserstein metric and stochastic MPC. Section III presents the modeling of static and dynamic uncertainties and a wide range of operational constraints in smart grids. Section IV provides the tube-based uncertain formulation for static uncertainties, the stochastic MPC formulation of dynamic uncertainties, the unified WDR-MPC framework, the distributionally robust reformulation, and the scalable acceleration methods. 
Section V displays the outcomes of numerical experiments conducted on actual distribution networks. Section VI concludes this paper.  


\section{Preliminaries}

\subsection{Wasserstein Metric}
In order to define a metric-based ambiguity set, we first need to define a metric between distributions. The Wasserstein metric used in this paper is defined as follows: 
\begin{mydef} \label{Wasserstein Metric}
    (Wasserstein metric \cite{mohajerin2018data}): The Wasserstein metric $d_{W}(\mathbb{P}_1,\mathbb{P}_2)$:$\mathcal{M}(\Xi)\times\mathcal{M}(\Xi) \longrightarrow \mathbb{R}$ is defined via
    \begin{equation} \nonumber
        d_{W}(\mathbb{P}_1,\mathbb{P}_2) = \inf_{\pi\in \Pi(\mathbb{P}_1,\mathbb{P}_2)}\left\{ \int_{\Xi^2}\Vert \xi_1-\xi_2 \Vert\pi(d\xi_1,d\xi_2) \right\},
    \end{equation}
    where $\Pi(\mathbb{P}_1,\mathbb{P}_2)$ is the set of joint probability distributions of $\xi_1,\xi_2$ with a marginal distribution of $\mathbb{P}_1$ and $\mathbb{P}_2$. $\ell_1$-norm is used here due to its superior numerical properties. 
\end{mydef} 

Given a series of observed samples $\{\Hat{\xi}_1,...,\Hat{\xi}_{N_s}\}$, where $N_s$ is the sample size, the empirical distribution can be determined from
\begin{equation} \label{empirical distribution}
    \Hat{\mathbb{P}}_{N_s} = \frac{1}{N_s} \sum_{k=1}^{N_s} \delta_{\Hat{\xi}_k},
\end{equation}
where $\delta_{\Hat{\xi}_k}$ is the Dirac distribution centered at $\Hat{\xi}_k$. Accordingly, the following ambiguity set can be constructed, centered at $\Hat{\mathbb{P}}_{N_s}$:
\begin{equation} \label{ambiguity set}
     \mathcal{B}_{\epsilon} = \left\{ \mathbb{P} \in \mathcal{M}(\Xi):d_{W}\left(\mathbb{P},\Hat{\mathbb{P}}_{N_s}\right) \leq \epsilon \right\},
\end{equation}
where $\epsilon$ is the radius of the ambiguity set, which controls how conservative the decisions are. 
Wasserstein ambiguity sets exhibit better statistical characteristics when contrasted with other types, leading to strong out-of-sample performance \cite{mohajerin2018data}.

\subsection{Stochastic Model Predictive Control}
Consider a linear discrete-time system with random disturbance: 
\begin{equation} \label{linear discrete-time MPC}
    \bm{x}(t+1) = A\bm{x}(t)+B\bm{u}(t)+\bm{w}(t),
\end{equation}
where $t$ is the time slot, $\bm{x}(t)$ is the state, $\bm{u}(t)$ is the control input, and the disturbance $\bm{w}(t)$ is generated from a (possibly unknown) distribution. Inspired by the idea of a tube-based MPC \cite{mayne2005robust}, the system can be partitioned into two parts: a nominal component $\bm{z}(t)$ and a cumulative error component $\bm{e}(t)$, i.e. $\bm{x}(t) = \bm{z}(t) + \bm{e}(t)$. 
The control feedback control policy is expressed as $\bm{u}(t) = \bm{v}(t) + K\bm{e}(t)$, where the feedback gain matrix $K$ is employed to ensure system stability and mitigate the impact of disturbances. A standard method of constructing $K$ is to solve a convex optimization problem involving a series of linear matrix inequalities. Further details can be found in Section 7.2 of Limon et al \cite{limon2010robust}.) 
After obtaining the gain matrix $K$, the stochastic MPC system is listed as follows: 
\begin{subequations} \label{MPC original}
\begin{align}
\min_{\bm{v},\bm{z}} \qquad & \bm{L}(\bm{z},\bm{v}) \\
\mathrm{s.t.} \qquad & \bm{z}(t+1) = A\bm{z}(t)+B\bm{v}(t) \\
& \bm{u}(t) = \bm{v}(t) + K\bm{e}(t) \\
& \bm{x}(t) = \bm{z}(t) + \bm{e}(t) \\
& \bm{e}(t+1) = (A+BK)\bm{e}(t) + \bm{w}(t), \\
& \bm{x} \in \bigotimes_{t=1}^{T} \mathbb{X}(t), \bm{u} \in \bigotimes_{t=1}^{T} \mathbb{U}(t), 
\end{align}
\end{subequations}
where $\bm{L}(\bm{z},\bm{v})$ is the MPC cost function, and $\mathbb{X}(t)$ and $\mathbb{U}(t)$ represent polytopes that define the feasible sets for states and inputs at each time point, respectively. 

\section{Problem formulation}
We consider a distribution power network that includes normal generators, RESs, and charging stations with PEVs. Our optimization problem involves both static and dynamic uncertainties in a distribution network within a stochastic MPC framework. The static uncertainties come from RESs, flexible loads, time-variant electricity prices, etc. 
Without loss of generality, our model formulation only considers the uncertain RESs as static uncertainties and uncertain aggregated charging/discharging power in charging stations as dynamic uncertainties. Other static uncertainties, such as flexible loads and time-variant electricity prices, could be similarly integrated into the model.

\subsection{Distribution Networks with Static Uncertainties}
This subsection outlines the stochastic formulation of static uncertainties of RESs and the operational constraints in the distribution network. We follow standard formulations to establish the model of a distribution network \cite{yeh2012adaptive,li2022distributionally}. Consider a radial distribution network characterized by a set of buses denoted as $\mathcal{N}:= {1, \ldots, N}$. The set $\mathcal{C}_k$ signifies the child nodes associated with bus $k \in \mathcal{N}$, and $\mathcal{A}_k$ represents the parent node of $k$. The connectivity within the distribution network is established by a set of power lines $\mathcal{L} \in \mathbb{R}^{N-1}$. A line $k \in \mathcal{L}$ denotes the connection between node $\mathcal{A}_k$ and node $k$. Denote $\mathcal{G} \subseteq \mathcal{N}$ $\mathcal{C} \subseteq \mathcal{N}$ as the sets of buses linked to distributed generators charging stations, respectively. 

Considering that there are forecasting errors associated with RESs, the power outputs are as follows: 
\begin{equation}
    \bm{\Tilde{P}}^R_k(t) = P^R_k(t) + \bm{\xi}_k(t),  \label{RES output}
\end{equation}
where $\bm{\xi}_k(t)$ denotes the forecasting error, $\bm{\Tilde{P}}^R_k(t)$ represents the actual power output of RESs, and $P^R_k(t)$ denotes the forecast of actual power output of the RESs, which is equal to the average output from historical data. Notably, if the node $k$ is not linked to any RESs, $\bm{\xi}_k(t),\bm{\Tilde{P}}^R_k(t)$ and $P^R_k(t)$ will be assigned  as 0. 

The conventional generators are presumed to adapt the power outputs using subsequent linear decision rules, aiming at alleviating the prediction discrepancies throughout the entire distribution network: 
\begin{equation} \label{generator output}
    \Tilde{\bm{P}}_{k}^{G}(t) = \bm{P}_{k}^{G}(t) -\bm{\alpha}_{k}(t)e^\top\bm{\xi}_k(t),\ k\in\mathcal{G}, 
\end{equation}
\begin{equation}\label{alpha bounds}
    \sum_{k\in\mathcal{N}}\bm{\alpha}_k(t) = 1, \ 0 \leq \bm{\alpha}_k(t) \leq 1,\ k\in\mathcal{N}
\end{equation}
\begin{equation}\label{generator active bounds}
    \underline{P}_k^G \leq \Tilde{\bm{P}}_k^G(t) \leq \overline{P}_k^G,\ k\in\mathcal{G}, 
\end{equation}
\begin{equation}\label{generator reactive bounds}
    \underline{Q}_k^G \leq \bm{Q}_k^G(t) \leq \overline{Q}_k^G, \ k\in\mathcal{G}, 
\end{equation}
where $e = [1,..,1]^\top \in \mathbb{R}^N$. Constraint \eqref{generator output} implies that the forecasting errors will be distributed across each generator, thereby causing fluctuations of active power output $\Tilde{\bm{P}}_k^G(t)$ by the generators. Constraint \eqref{alpha bounds} ensures that the forecasting errors are fully mitigated. Constraints \eqref{generator active bounds} and \eqref{generator reactive bounds} guarantee that the generators' active and reactive power outputs fall within their physical bounds, respectively.  In cases where node $k$ has no connection to a controllable generator, $\bm{\alpha}_k(t)$, $\bm{P}^G_k$, and their respective bounds are all assigned as 0.

Given the radial configuration of the analyzed distribution system, each line carries the complex load of all nodes. In this context, $\bm{P}_k(t) \in \mathbb{R}$ represents the power flow along the line terminating at node $k$. Here we employ the linearized DistFlow approximation for AC power flow as described in \cite{yeh2012adaptive}. The power flow along the distribution line impacted by uncertainty is listed below: 
\begin{equation} \label{line capacity}
    \Tilde{\bm{P}}_k(t) = \bm{P}_k(t) -B_{i*}\left(\bm{\xi}(t) -\bm{\alpha}(t) e^\top\bm{\xi}(t)\right),
\end{equation}
where $B_{i*}$ denotes the $i$-th row of $B\in\mathbb{R}^{|\mathcal{L}|\times|\mathcal{N}|}$ where its entries are defined as $B_{ij} = 1$ if line $i$ is part of the path from the root to bus $j$, and $B_{ij} = 0$ otherwise. In the context of a distribution network, the voltage at node $k$ can be determined from the active power $\bm{P}_k$, reactive power $\bm{Q}_k$, and the voltage at its parent node $\mathcal{A}_k$, as follows:
\begin{equation}
    \bm{V}_k(t) = \bm{V}_{\mathcal{A}_k}(t)-(r_k\bm{P}_k(t)+x_k\bm{Q}_k(t))/V_0,\label{volatage1}
\end{equation}
where $r_k$ and $x_k$ respectively represent the real and imaginary resistance of line $k$, and $V_0$ denotes the basic voltage. Subsequently, the voltage at each node influenced by uncertainty is deduced from \eqref{volatage1}, leading to 
\begin{equation}
    \begin{aligned}\label{volatage2}
    \Tilde{\bm{V}}_k(t) &= \bm{V}_{\mathcal{A}_k}(t)-\left(r_k\Tilde{\bm{P}}_k(t)+x_k\Tilde{\bm{Q}}_k(t)\right)/V_0\\
                    &= \bm{V}_k(t)+B_{*i}^\top\left[RB\left(\bm{\xi}(t) -\bm{\alpha}(t) e^\top\bm{\xi}(t)\right)\right]/V_0,
    \end{aligned}
\end{equation}
where $R$ is the square matrix of size $N\times N$, where the diagonal entries are composed of the line resistances ($R_{ii}=r_i$), and all off-diagonal entries are zero ($R_{ij}=0$ for $i\neq j$). The voltage at each node needs to adhere to the following bounds: 
\begin{equation}
    \underline{V}_k \leq \Tilde{\bm{V}}_k(t) \leq \overline{V}_k.
    \label{volatage constraints}
\end{equation}

\vspace{-0.4cm}
\subsection{Charging Stations with Dynamic Uncertainties}
This subsection introduces the settings of the stochastic MPC framework that governs charging stations with dynamic uncertainties. Since the arrival and departure time of PEVs are generally not known in advance, it is important to consider the uncertainty of PEV charging/discharging behavior. Many existing studies \cite{V2G1,V2G2,zeynali2020multi,wang2019joint,nguyen2023distributionally} have assumed that the number of PEVs at charging stations is known beforehand, which is not realistic in most practical situations. 
\textcolor{black}{In our previous study \cite{li2022distributionally}, we employed an MPC method eliminating the need for prior knowledge of PEV numbers. However, a limiting assumption was made that some EVs would discharge to meet others' charging needs, deviating from real-world scenarios where frequent charging/discharging can harm battery cycling age.}

\textcolor{black}{To solve the reliability issues, charging stations with large batteries are introduced. The battery serves as the reserve for the uncertain charging demands. Based on this assumption, the battery's real charging/discharging power $\Tilde{\bm{P}}^{B}_{k}(t)$ suffers from uncertainties:
\begin{equation}
    \Tilde{\bm{P}}^{B}_{k}(t) =  \bm{P}^{B}_{k}(t) + \bm{\omega}_k(t).
\end{equation}
where $\bm{P}^{B}_{k}(t)$ is the estimated charging power, and $\bm{\omega}(t)$ at each time slot $t$ is the error between forecast charging demand and real charging demand. The charging dynamic is listed below:}
\begin{equation} \label{MPC update rules}
    \bm{SoC}_{k}(t+1) = \bm{SoC}_{k}(t) + \bm{P}^{B}_{k}(t) + \bm{\omega}_k(t),
\end{equation}
where $t \in \mathcal{T}$, $\mathcal{T} = \left\{t_0,...,t_0+T_c-1\right\}$ denotes the MPC period and $T_c$ is the MPC time range. 
\textcolor{black}{The uncertainty $\bm{\omega}_k(t)$ is regarded as corresponding to dynamic uncertainties instead of static uncertainties like RES forecast errors, because these uncertainties may accumulate through the dynamic system and impact the future states. Specifically, we illustrate the difference through a simple example by considering the SoC at time slot $t+2$:
\begin{subequations}
    \begin{align}
        \bm{SoC}_{k}(t+2) &= \bm{SoC}_{k}(t+1) + \bm{P}^{B}_{k}(t+1) + \bm{\omega}_k(t+1), \nonumber\\
                          &= \bm{SoC}_{k}(t) + \bm{P}^{B}_{k}(t) + \bm{P}^{B}_{k}(t+1) \nonumber\\
                          & \qquad \qquad \qquad \qquad + \bm{\omega}_{k}(t) + \bm{\omega}_{k}(t+1). \nonumber
    \end{align}
\end{subequations}
As we can see in the above equation, the random disturbance $\bm{\omega}_{k}(t)$ could still influence future states and the influences accumulate as time goes by. This, of course, creates additional difficulties when designing distributionally robust optimization methods.}

Some standard physical constraints of $\bm{SoC}_{k}(t)$ and $\bm{P}^{B}_{k}(t)$ are as follows: 
\begin{equation}\label{battery charging constraints}
    -\overline{P}_{dch,k} \leq \bm{P}_{k}^B(t) \leq \overline{P}_{ch,k},
\end{equation}
\begin{equation}
     \underline{C}_k \leq \bm{SoC}_k(t) \leq \overline{C}_k, \label{SoC constraints}
\end{equation}
where $\overline{P}_{dch,k}$ and $\overline{P}_{ch,k}$ represent the maximum charging/discharging power of batteries; $\underline{C}_k$ and $\overline{C}_k$ denote the lower bounds and upper bounds of the battery capacity, respectively. When node $k$ is not linked to charging stations, the variables are assigned as 0. 
\textcolor{black}{$P^{Cpred}_k(t)$ represents the predicted total charging demand of the charging station at node $k$. $\bm{P}^{CS}_k(t)$ denotes the total power transmitted to the charging station at node $k$  through the distribution network:
 \begin{equation}
     \bm{P}^{CS}_k(t) = P^{Cpred}_k(t) + \bm{P}^{B}_{k}(t). \label{charging stations output}
 \end{equation}
}

\vspace{-0.4cm}
\subsection{Overall Optimization Formulation}
This subsection begins with the overall formulations of the constraints and operational costs. Then the optimization problem is set out, including its uncertain constraints and deterministic constraints. It is imperative to consider various factors, including the incoming power flow from the parent node, the outgoing power flow to child nodes, power consumption, and power generation, etc. The power consumption encompasses both the active loads $P_k^L(t)$ and inactive loads $Q_k^L(t)$ of node $k$, along with the charging load from PEVs $\bm{P}_k^B(t)$. Consequently, the balance of active/inactive power flow at node $k$ can be expressed:  
\begin{equation}\label{active power balance}
    \bm{P}_k(t) = \sum_{j\in\mathcal{C}_k}\bm{P}_j(t)-\bm{P}_k^G(t)-P_k^R(t)+\bm{P}_k^{CS}(t)+P_k^L(t),
\end{equation}
\begin{equation}\label{reacitve power balance}
    \bm{Q}_k(t) = \sum_{j\in\mathcal{C}_k}\bm{Q}_j(t)-\bm{Q}_k^G(t)+Q_k^L(t).
\end{equation}
Since electricity prices fluctuate, the battery can efficiently exploit this variability by charging during periods of low prices and discharging during periods of high prices. This can be quantified through capacity loss, often assessed by means of cycling age \cite{wang2014degradation}. For our computations, we chose a cycling age model tied to cumulative throughput, as formulated by Antoniadou-Plytaria et al.\cite{antoniadou2020market}: 
\begin{equation} \label{degradation cost1}
    \bm{T}_{k} = B_1e^{B_2I_c}\sum_{t=t_0}^{t_0+T_c-1} \left\vert \bm{P}^B_{k}(t)\right\vert \Delta t,
\end{equation}
where $\bm{T}_{k}$ denotes the proportion of battery capacity loss, while $B_1$ and $B_2$ are coefficients derived from empirical observations, and $\Delta t$ stands for the time interval. Accordingly, we obtain the following PEV battery cost:
\begin{equation} \label{degradation cost2}
    \bm{c}_{k} = \frac{C_{k}^B\bm{T}_{k}}{1-\eta_{end}},
\end{equation}
where $C_{k}^B$ represents the cost of battery installation at the charging station of node $k$ and $\eta_{end}$ is the remaining capacity of the battery life in percentage terms. Thus, the total operational cost can be determined by the following cost function: 
\begin{align}\label{cost function}
    \bm{J}_{cost}=&\sum_{t=t_0}^{t_0+T_c-1} \sum_{k \in \mathcal{N}}[f_g\left(\bm{P}^G_{k}(t) \right) \nonumber \\ &\qquad\qquad\qquad\qquad + c_{B}(t)\bm{P}_{k}^B(t)]+\sum_{k\in \mathcal{B}}\bm{c}_{k}, 
\end{align} 
where $c_{B}(t)$ denotes charging price at time slot $t$. Here, the control variables can be collected into $\bm{y}(t) = [\bm{P}^G(t),\bm{Q}^G(t),\bm{\alpha}(t),\bm{V}(t),\bm{P}(t), \bm{Q}(t), \bm{P}^B(t), \bm{SoC}(t)]^\top$, and the overall MPC formulation becomes
\begin{subequations} \label{OPT}
    \begin{align}
    \min_{\bm{y}(t)} &  \qquad \bm{J}_{cost} \label{OPT obj}\\
    \mathrm{s.t.} & \qquad \eqref{alpha bounds},\eqref{generator reactive bounds},\eqref{charging stations output},\eqref{active power balance},\eqref{reacitve power balance},\eqref{degradation cost1} \label{certain constraints},\\
                  & \qquad \eqref{generator output},\eqref{generator active bounds},\eqref{volatage2},\eqref{volatage constraints},\label{uncertain constraints}\\
                  & \qquad \eqref{MPC update rules}-\eqref{SoC constraints}, \label{MPC uncertain rules}
    \end{align}
\end{subequations}
where \eqref{certain constraints} collects the deterministic constraints without any uncertainties, \eqref{uncertain constraints} collects the constraints with static uncertainties and \eqref{MPC uncertain rules} collects the constraints with dynamic uncertainties of MPC. Constraint \eqref{uncertain constraints} is discussed further in Section IV. A, while constraint \eqref{MPC uncertain rules} is discussed in Section IV. B.

\section{Solution Approach}
Given that uncertainty can be either static or dynamic, finding a feasible solution for Eq. \eqref{OPT} can be quite challenging. Inspired by tube-based techniques in MPC frameworks, we sought to formulate a tube that would be invariant to uncertainty. 
Ultimately, we devised two different approaches for formulating distributionally-robust invariant tubes using Eq. \eqref{uncertain constraints} and Eq. \eqref{MPC uncertain rules}. This leads to a tractable convex reformulation and a method of accelerating the WDR optimization. This section concludes with a summary of the WDR-MPC framework.

\vspace{-3mm}
\subsection{Static Uncertainty Formulation}
In terms of the uncertain constraints \eqref{uncertain constraints}, the random variables only affect decision variables in the current time slot $t$. Hence, a multi-objective loss function needs to be optimized that includes the operational cost functions as well as the risk functions of the random variables. Thus, Eq. \eqref{uncertain constraints} needs to be reformulated as 
follows:
\begin{subequations}
\begin{align}
    \underline{P}^G_k \leq & \bm{P}_k^G(t) -\bm{d}_k^G(t) \ , \text{and} \ \bm{P}_k^G(t) + \bm{u}_k^G(t) \leq \overline{P}_k^G, \label{PG constraints}\\
    \underline{V}_k \leq & \bm{V}_k(t) - \bm{d}_k^V(t) \ , \text{and} \ \bm{V}_k(t) + \bm{u}_k^V(t) \leq \overline{V}_k, \label{V constraints}\\
    & -\bm{d}_k^G(t) \leq -\bm{\alpha}_k(t)e^\top \bm{\xi}_k(t) \leq \bm{u}_k^G(t), \label{dGuG}\\
    & -\bm{d}_k^V(t) \leq \bm{\mu}_k(t)^\top \bm{\xi}_k(t) \leq \bm{u}_k^V(t), \label{dVuV}
\end{align}
\end{subequations}
where $\bm{\mu}_k(t):={B}_{*k}^\top{R}{B}\left({I}-\bm{\alpha}(t){e}^\top\right)^\top/V_0$ is introduced to simplify the formulation,  $\bm{d}_k^{G/V}(t)$ and $\bm{u}_k^{G/V}(t)$ are auxiliary optimization variables. 
Following a standard CVaR procedure \cite{rockafellar2000optimization}, we can transform the uncertain constraints \eqref{dGuG} and \eqref{dVuV} into risk functions, leading to the following proposition: 
\begin{mypro} \label{CVaR Cor}
    The risk function $\bm{J}_{risk}^G$ and $\bm{J}_{risk}^V$ for \eqref{dGuG} and \eqref{dVuV} are respectively,
    \begin{subequations}
    \begin{equation} \label{riskG}
    \bm{J}_{risk}^G = \sum_{t=t_0}^{t_0+T_c-1} \sum_{k \in \mathcal{N}} \mathbb{E}^{\mathbb{P}} \left\{\max_{j\in\{1,2,3,4\}}\left[ \la \bm{a}^G_{jk},\bm{\xi}\ra + \bm{b}_{jk}^G\right]\right\},
    \end{equation}
    \begin{equation} \label{riskV}
    \bm{J}_{risk}^V = \sum_{t=t_0}^{t_0+T_c-1} \sum_{k \in \mathcal{N}} \mathbb{E}^{\mathbb{P}} \left\{\max_{j\in\{1,2,3,4\}}\left[ \la \bm{a}^G_{jk},\bm{\xi}\ra + \bm{b}_{jk}^G\right]\right\}.
    \end{equation}
    \end{subequations}
\end{mypro}
A detailed derivation is provided in Appendix A. For simplicity, we let $\bm{J}_{risk} = \bm{J}_{risk}^V + \bm{J}_{risk}^G$, $\bm{J}_{norm} = \vert \bm{d}_k^G + \bm{u}_k^G\vert +\vert \bm{d}_k^V + \bm{u}_k^V\vert$. Thus, we have a new optimization problem:
\begin{subequations}\label{OPT2}
    \begin{align} 
        \min_{\bm{y}(t),\bm{\gamma}(t)} &  \qquad \bm{J}_{cost} + \mu_1\sup_{\mathbb{P}} \bm{J}_{risk} + \mu_2 \bm{J}_{norm}\\
    \mathrm{s.t.} & \qquad \eqref{certain constraints}, \eqref{PG constraints}, \eqref{V constraints},  \eqref{MPC uncertain rules}, \label{OPT2a}
    \end{align}
\end{subequations}
where $\mu_1$ and $\mu_2$ are the weighting factors between the cost functions and risk functions, and $\bm{y}(t)$ is defined in the formulation \eqref{OPT}. All auxiliary variables are collected in $\bm{\gamma}(t) = [\bm{d}_k^{V/G},\bm{u}_k^{V/G},\bm{\omega}_{1/2}^{V/G}]^\top$.

\subsection{Stochastic MPC formulation of dynamic uncertainties}
With respect to the dynamic uncertainties in Eq. \eqref{MPC uncertain rules}, a tube-based MPC is used to construct the ambiguity tube. From the MPC constraints in \eqref{MPC uncertain rules}, one can see that our MPC model is a linear discrete-time system of a similar form \eqref{linear discrete-time MPC}, where $\bm{x}$ stands for $\bm{SoC}(t)$, and $\bm{u}$ stands for $\bm{P}^B(t)$. The other variables in $\bm{y}(t)$ do not affect the MPC dynamics. Here we follow Ref. \cite{fochesato2022data} to define the ambiguity tube $\mathbb{S}(t)$ for error $\bm{e}(t)$. Here, $\mathbb{S}(t)$ is restricted to a hyper-rectangular shape with lower bound set $\underline{\bm{\alpha}} = \{ \{\overline{\bm{\alpha}}_{k}^{j}\}_{k=1}^{T}\}_{j=1}^{d}$  and upper bound set $\overline{\bm{\alpha}} = \{\{\overline{\bm{\alpha}}_{k}^{j}\}_{k=1}^{T}\}_{j=1}^{d}$, where $d$ is the dimension of $e(t)$. 
Then, the ambiguity tubes $\mathbb{Z}(t)$ and $\mathbb{V}(t)$ are constructed accordingly 
\cite{farina2016stochastic}: 
\begin{align}
    \mathbb{Z}(t) = \mathbb{X}(t) \ominus \mathbb{S}(t), t \in \mathcal{T}, \label{feasible region Z}\\
    \mathbb{V}(t) = \mathbb{U}(t) \ominus K\mathbb{S}(t), t \in \mathcal{T} \label{feasible region V},
 \end{align}
where $\ominus$ represents the operator of the Pontryagin set difference. After decoupling the uncertainties via ambiguity tubes, we have a normal MPC system: 
\begin{subequations} \label{SMPC}
\begin{align}
    \min_{\bm{v},\bm{z}} \qquad & \bm{L}(\bm{z},\bm{v}) \\
    \mathrm{s.t.} \qquad & \bm{v} \in \bigotimes_{t=1}^{T} \mathbb{V}(t), \bm{z} \in \bigotimes_{t=1}^{T} \mathbb{Z}(t), \\
     & \bm{z}(t+1) = A\bm{z}(t)+B\bm{v}(t).
\end{align}
\end{subequations}

According to the formulation \eqref{SMPC}, the MPC optimization problem could be solved in two stages. Therefore, in this subsection, our focus is on deriving some distributionally-robust bounds for the ambiguity tube $\mathbb{S}(t)$ of $\bm{e}(t)$. 

Assume we are given a dataset $\Hat{\mathcal{D}} = \left\{\Hat{w}_i\right\}_{i=1}^{n}$ containing $nT$-long i.i.d. disturbance sequences, where $\Hat{\bm{w}}_i = \left[\Hat{w}_i(t_0)^\top,...,\Hat{w}_i(t_0+T_c-1)^\top\right]^\top \in \mathbb{R}^{dT}$. A corresponding $e(t)$ can be obtained from a historical dataset of $\bm{w}(t)$, where $\Hat{e}_i(t) = \sum_{j=0}^{k-1}(A+BK)^{k-1-j}\Hat{w}_i(j)$ and $\Hat{e}_i(0) = 0$. Accordingly, the following optimization problem is solved to establish the ambiguity tube: 
\begin{subequations}\label{ambiguous tube optimization}
    \begin{align}
\min_{\underline{\bm{\alpha}}_{k}^{j},\overline{\bm{\alpha}}_{k}^{j}} \qquad & \overline{\bm{\alpha}}_{k}^{j}-\underline{\bm{\alpha}}_{k}^{j}, \\
    \mathrm{s.t.} \qquad & \underline{\bm{\alpha}}_{k}^{j} \leq \overline{\bm{\alpha}}_{k}^{j},  \label{ambiguous tube certain} \\
                         & \underline{\bm{\alpha}}_k^j \leq \bm{e}^j(k) \leq \overline{\bm{\alpha}}_k^j,\label{ambiguous tube uncertain}
    \end{align}
\end{subequations} 
where $\bm{e}^j(k)$ is the $j$-th dimension of $\bm{e}(k)$. Since $\bm{e}^j(k)$ is a random variable here, constraint \eqref{ambiguous tube uncertain} involves uncertainty. Hence \textbf{Proposition} \ref{CVaR Cor} is used to derive the risk functions and corresponding distributionally robust optimization problem: 
\begin{subequations}\label{bounds_CVaR}
    \begin{align} 
        \min_{\bm{y}_1} \qquad & \overline{\bm{\alpha}}_{k}^{j}-\underline{\bm{\alpha}}_{k}^{j} + \sup_{\mathbb{Q}} \mathbb{E}^{\mathbb{Q}}\left\{\bm{l}(\bm{y}_1,\bm{e}^{j}(k))\right\},\\
    \mathrm{s.t.} \qquad & \underline{\bm{\alpha}}_{k}^{j} \leq \overline{\bm{\alpha}}_{k}^{j},
    \end{align}
\end{subequations}
where $\bm{y}_1 = [\underline{\bm{\alpha}}_{k}^{j},\overline{\bm{\alpha}}_{k}^{j},\bm{\omega}_1,\bm{\omega}_2]^\top$. The risk function is listed as below: $\bm{l}_{1}(\bm{y}_1,\bm{e}^{j}(k)) = \frac{\mu}{1-\beta}\left\{\left[-\bm{e}^{j}(k) +\underline{\bm{\alpha}}_{k}^{j}-\bm{\omega}_1\right]_{+}+\left[\bm{e}^{j}(k) -\overline{\bm{\alpha}}_{k}^{j}-\bm{\omega}_2\right]_{+}\right\}$, where $\mu$ and $\beta$ denotes the weights and confidence level. The derivation is similar to \textbf{Proposition} \ref{CVaR Cor} in Appendix A with some simple substitutions.
\vspace{-0.3cm}
\subsection{WDR optimization reformulation and acceleration} 
This subsection introduces the distributionally robust formulation of MPC based on the Wasserstein metric, the corresponding tractable convex reformulation, and the acceleration method. 

Since the underlying distribution $\mathbb{P}$ and $\mathbb{Q}$ used in \eqref{OPT2} and \eqref{bounds_CVaR} may not be known beforehand in real-time situations, it can sometimes be quite difficult to optimize over an unknown distribution. For this reason, we have formulated the problem into a WDR optimization problem of the following form: 
\begin{subequations} \label{DRO}
    \begin{align}
        \min_{\bm{x}} \quad & \quad \sup_{\mathbb{P} \in \mathcal{B}_{\epsilon}}  \mathbb{E}^{\mathbb{P}} \left[\bm{l}(\bm{x},\xi) \right],\\
    \mathrm{s.t.} \quad & \quad \bm{x} \in \mathcal{X}, 
    \end{align}
\end{subequations}
where $\bm{x}$ is the control variable and $\mathcal{X}$ is the convex feasible region of $\bm{x}$, $\mathcal{B}_{\epsilon}$ is a Wasserstein ball following the definition \eqref{ambiguity set}. The distributionally-robust optimization has now been transformed into a min-max optimization problem. However, the inner optimization problem is an infinite-dimensional optimization problem, which is quite challenging to solve via current methods. Motivated by recent advances in WDR optimization \cite{mohajerin2018data}, we develop a tractable convex reformulation for the inner optimization problem in \eqref{DRO} as follows: 
\begin{subequations} \label{DRO reformulation}
    \begin{align}
    \inf_{\substack{\bm{\lambda},\bm{s}_i,\bm{z}_{i}, \bm{\nu}_{i},\bm{x}}}  & \bm{\lambda}\epsilon+\frac{1}{N_s} \sum_{i=1}^{N} \bm{s}_i ,\\
     \mathrm{s.t.} \quad & [\bm{l}]^{*}(\bm{z}_{i}-\bm{\nu}_{i}) + \sigma_{\Xi}(\bm{\mu}_{i})-\la \bm{z}_{i},\Hat{\xi}_i\ra \leq \bm{s}_i \label{DRO reformulation constraints1}, \\
     & \Vert \bm{z}_i \Vert_{\infty} \leq \lambda, \forall i \in \{1,...,N_s\}, \label{DRO reformulation constraints2} \\
     & \bm{x} \in \mathcal{X} ,
    \end{align}
\end{subequations}
where $\Xi$ is the domain of random variable $\bm{\xi}$, i.e., $\left[\underline{\xi}; \overline{\xi}\right]$, $[-\bm{l}]^{*}$ is the conjugate function of $-\bm{l}$ and $\sigma_{\Xi}$ is the support function of $\bm{\Xi}$. However, we can see that the number of constraints will increase as the sample size grows, which could ultimately lead to a heavy computation burden. To alleviate this issue, we develop an upper approximation that is more scalable to sample size compared to Eq. (\ref{DRO reformulation}).

\begin{mypro} \label{DRO reformulation cor}
    The optimization problem \eqref{DRO} can be approximated by the following upper approximation:
    \begin{subequations} \label{acce DRO}
        \begin{align}
            \inf_{\bm{\lambda},\bm{x}} \qquad & \bm{\lambda}\epsilon+\frac{1}{N}\sum_{i=1}^{N_s} \bm{l}(\bm{x},\Hat{\xi}_i), \\
            \mathrm{s.t.} \qquad & \bm{\lambda} \geq \lambda_0 \quad \text{and} \quad \bm{\lambda} \geq 0, \\
                            & \bm{x} \in \mathcal{X},
        \end{align}
    \end{subequations}
    where $\lambda_0 = \max\{\eta(\bm{x},\underline{\xi}),\eta(\bm{x},\overline{\xi})\}$ and $\eta(\bm{y},\xi) = \frac{\partial \bm{l}(\bm{x},\xi)}{\partial \xi}$.
\end{mypro}
A detailed derivation is given in Appendix B. Note that the number of optimization variables and constraints does not grow with the sample size. Therefore, \eqref{acce DRO} has a much more favorable computation overhead. In fact, the gap between \eqref{DRO reformulation} and \eqref{acce DRO} is very small and decreases as the sample size increases.

In summary, what we have outlined here is an accelerated WDR optimization solution that is scalable to the sample size. It is also a finite-dimensional tractable convex optimization problem that can be solved directly by mature solvers. 
\vspace{-3mm}
\subsection{Summary of WDR-MPC}
In this subsection, we combine the formalisms of the preceding three subsections and construct a unified WDR-MPC framework for handling both static and dynamic uncertainties.

The first step is to reformulate the optimization problem \eqref{bounds_CVaR} based on the acceleration technique in \eqref{acce DRO}: 
\begin{subequations} \label{DRMPC stage1}
    \begin{align}
        \min_{\bm{y}_1,\bm{\lambda}} \qquad 
        & \overline{\bm{\alpha}}_k^j - \underline{\bm{\alpha}}_k^j + \bm{\lambda}\epsilon + \sum_{i=1}^{N_s} \bm{l}_1(\bm{y}_1,\Hat{\bm{e}}^j(k)), \\
        \mathrm{s.t.} \qquad & \underline{\bm{\alpha}}_{k}^{j} \leq \overline{\bm{\alpha}}_{k}^{j}, \\
                             & \bm{\lambda} \geq \lambda_1 \quad \text{and} \quad \bm{\lambda} \geq 0,
    \end{align}
\end{subequations}
where $\lambda_1 = \max\{\bm{l}_1'(\bm{y}_1,\underline{e_k^j}),\bm{l}_1'(\bm{y}_1,\overline{e_k^j})\}$ and $\bm{l}_1'(\bm{y}_1,e_k^j) = \frac{\partial \bm{l}_1(\bm{y}_1,e_k^j)}{\partial e_k^j}$. Supposing the optimal solution of optimization problem \eqref{DRMPC stage1} as $\underline{\alpha}_t^{j,*}$ and $\overline{\alpha}_t^{j,*}$,  $\mathbb{S}(t)$ in time slot $t$ can be derived by the following formulation:
\begin{equation}
    \mathbb{S}(t) = \bigotimes_{j=1}^{d}\left[\underline{\alpha}_{t}^{j,*},\overline{\alpha}_{t}^{j,*}\right].
\end{equation}
Equations \eqref{feasible region Z} and \eqref{feasible region V} then provide the ambiguity tubes $\mathbb{Z}(t)$ and $\mathbb{V}(t)$ and the stochastic MPC optimization problem \eqref{OPT2} is reformulated into a normal MPC problem:
\begin{subequations}\label{MPC stage2}
    \begin{align}
        \min_{\bm{y}} \qquad & \bm{l}(\bm{y},\xi) \\
    \mathrm{s.t.} \qquad & \bm{SoC}_{k}(t+1) = \bm{SoC}_{k}(t)+\bm{P}^B_k(t) \label{MPC stage2a}\\
                         & \bm{SoC}_k(t) \in \mathbb{Z}_k(t), t \in \mathcal{T}, \label{MPC stage2b} \\
                         & \bm{P}^B_k(t) \in \mathbb{V}_k(t), t \in \mathcal{T}, \label{MPC stage2c}\\
                         & \eqref{certain constraints},\eqref{PG constraints},\eqref{V constraints},\label{MPC stage2d}
    \end{align}
\end{subequations}
where $\bm{l}(\bm{y},\xi)$ is the optimization objective function in \eqref{OPT2} and
\begingroup
\footnotesize
$\bm{y}(t) = [\bm{P}^G(t),\bm{Q}^G(t),\bm{\alpha}(t),\bm{V}(t), ,\bm{P}(t),\bm{Q}(t), \bm{P}^B(t),\bm{SoC}(t)]^\top.$
\endgroup

After transferring influences created by the dynamic uncertainty through the ambiguity tubes, the next priority is to reformulate the WDR optimization for static uncertainty in each time slot $t$. \textbf{Proposition} 2 is also used here so as to handle the worst-case risk costs $\sup_{\mathbb{P}}\bm{l}(\bm{y},\xi)$. Accordingly, the overall WDR-MPC is reformulated into the following finite-dimensional tractable convex optimization problem:
\begin{subequations} \label{DRMPC stage2}
    \begin{align}
        \min_{\bm{y},\bm{\gamma},\bm{\lambda}} \qquad & \bm{\lambda}\epsilon + \sum_{i=1}^{N_s}\bm{l}(\bm{y},\Hat{\xi}_i), \\
        \mathrm{s.t.} \qquad & \eqref{MPC stage2a}-\eqref{MPC stage2d}, 
    \end{align}
\end{subequations}
with $\bm{\gamma}(t) = [\bm{d}_k^{V/G},\bm{u}_k^{V/G},\bm{\omega}_{1/2}^{V/G}]^\top$,  $\lambda_0 = \max\{\bm{l}'(\bm{y},\underline{\xi}),\bm{l}'(\bm{y},\overline{\xi})\}$, and $\bm{l}'(\bm{y},\xi) = \frac{\partial \bm{l}(\bm{y}_1,\xi)}{\partial \xi}$.

To sum up, this section provides a detailed procedural outline of our proposed WDR-MPC method, as summarized in the \textbf{Algorithm} 1: 
\begin{algorithm}[!h]
    \label{algorithm}
    \caption{WDR-MPC}
    \begin{algorithmic}
        \State \textbf{Require}: Historical disturbance $\Hat{\mathcal{D}}$, historical RESs data $\left\{\Hat{\bm{\xi}}_i\right\}$, MPC period number $T_l$, system settings and fixed hyperparameters.
        \State \textbf{Result}: Control actions:  $\bm{y}(t) = [\bm{P}^G(t),\bm{Q}^G(t),\bm{\alpha}(t),\bm{V}(t),\bm{P}(t), \bm{Q}(t), \bm{P}^B(t), \bm{SoC}(t)]^\top$.
        \State 1. Solve problem \eqref{DRMPC stage1} with $\Hat{\mathcal{D}}$ to obtain the distributionally robust bounds of $\mathbb{Z}(t)$ and $\mathbb{V}(t)$.
        \While {$t_0 \leq T_l$}
        \State 2. Solve the WDR optimization problem \eqref{DRMPC stage2} to get the optimal control actions.
        \State 3. Collect the first computed control action into the final control action and update the $\bm{SoC}_k(t)$.
        \State 4. $t_0 = t_0 + 1$.
        \EndWhile
    \end{algorithmic}
\end{algorithm}

\section{Numerical results}
This section describes numerical experiments conducted to validate the feasibility of our method. The section begins with the model settings, after which we show some numerical results on the IEEE 38-bus system followed by the IEEE 94-bus system.

The problems were solved by the MOSEK \cite{mosek} solver through the CVX\cite{grant2014cvx} optimization toolkit in MATLAB. The experiments were performed on a personal laptop with an Intel Core i5, 1.2 GHz with 4 cores and 8 GB RAM. In our setup, one time slot represents an hour. The distribution of the charging demand at the charging stations was set to be consistent with the realistic situation in which there are more PEVs in the network in the daytime and fewer at night. 

The MPC time horizon $T_C$ was 12 hours with a receding horizon $T_l$ of 12. The cost of battery installation was \$150/KWh, the battery capacity $\overline{C}_k$ of the charging station was 0.3 MWh and $\underline{C}_k$ was 0 MWh. The maximum charging power $\overline{P}_{ch,k}$ and discharging power $\underline{P}_{dch,k}$ were both set to 0.1 MWh. The charging/discharging efficiency $\eta_{end}$ was set to 0.8. The control gain matrix $K$ was derived by the method described in Section 7.2 of Limon et al. \cite{limon2010robust}. The disturbances $w(t)$ were randomly generated from an interval with the predicted charging demand at the center and a charging demand radius of $20\%$. The power demand without PEVs was selected from a dataset of electricity demand in New York on 29 June 2019 \cite{ruggles2020developing}. The charging electricity price was taken from online data from Finland on 22 June 2022 \cite{2022See}. 
The solar data was taken from Pulazza et al. \cite{pulazza2021transmission} and was scaled to suitable values in the simulation. Moreover, to ensure the distribution generators took responsibility for generating power across the distribution network, we increase the price of transmission generation.\textcolor{black}{} As the prediction method is main scope of our method, we simply employed the mean value of the samples to estimate the predicted RES power generation, denoted as $P^{R}(t)$. 

As mentioned, we performed numerical experiments with the IEEE 38-bus and IEEE 94-bus systems as the power grid models with structure, physical limits, and cost function $f_g\left(P_{k}^{G}(t)\right)$ from Matpower \cite{zimmerman2010matpower}. Further details of the settings used are given in each subsection. 

\subsection{IEEE 38-bus system}
The IEEE 38-bus system comprises 38 buses and 37 lines. Node 1 is connected to the transmission system. The distribution network is modified by incorporating 3 distribution generators (Nodes 2, 6, and 12), 5 RES generators (Nodes 6, 16, 20, 25, and 28), and a single charging station (Node 2). The values for $\underline{P}^{G}_{k}$, $\overline{P}^{G}_{k}$, $\underline{V}_{k}$, and $\overline{V}_{k}$ were established at 0 MW, 0.75 MW, 0.95 p.u., and 1.05 p.u. respectively. 

The simulation results for different Wasserstein radii $\epsilon$ are shown in Table \ref{different epi radius norm}.
\begin{table}[htbp!]
    \centering
    \caption{Tube radius of dynamic and static uncertainties with different $\epsilon$}         \renewcommand{\arraystretch}{1.5}
    \begin{tabular}{|c|c|c|c|c|c|c|}
    \hline
    $\epsilon$ & 0.000 & 0.001 & 0.002 & 0.005 \\
     \hline
     Dynamic Tube radius &0.8241& 0.8092 & 0.8058 & 0.8025 \\
     \hline
     Static Tube Radius & 11.20 & 11.15 & 9.92 & 9.92 \\
     \hline
    \end{tabular}
\label{different epi radius norm}
\end{table}
We can see from the table that when $\epsilon$ grows, the interval radius tends to shrink. But, as the Wasserstein radius increases, the variations in the random disturbances grow bigger. This indicates that the invariant set tends to be smaller, as shown by the tube radius.

Concerning the computational speed, we compared WDR-MPC and accelerated WDR-MPC with different sample sizes in Table \ref{time comparison}. Since the acceleration method is applied in two stages, we only conduct numerical experiments of the stage-2 optimization problem \eqref{DRMPC stage2} and present the average computation time of the WDR-MPC and accelerated WDR-MPC within time range $T_c = 16$ as we can see in Table \ref{time comparison}. WDR-MPC is well ahead of WDR-MPC in computational speed, and the speed gap tends to be larger when the sample size grows. Moreover, WDR-MPC without acceleration couldn't solve the problem in the 50000s, which exceeds the MPC time range. This further indicates the superiority of our accelerated WDR-MPC. 
\begin{table}[htbp!]
    \centering
    \caption{Time consumption of WDR-MPC and accelerated WDR-MPC}
    \renewcommand{\arraystretch}{1.5}
    \begin{tabular}{|c|c|c|c|c|}
    \hline
    sample size & 10 & 30 & 50 & 100\\
    \hline
    WDR-MPC(s) & 733.04 & 5143.64 & 13842.33 & 50000+ \\
    \hline
    accelerated WDR-MPC(s) & 135.96 & 463.04 & 896.63 & 2660.64  \\
    \hline
    \end{tabular}
\label{time comparison}
\end{table}
\begin{figure}[htbp!]
    \centering
    \includegraphics[width= 0.9 \columnwidth]{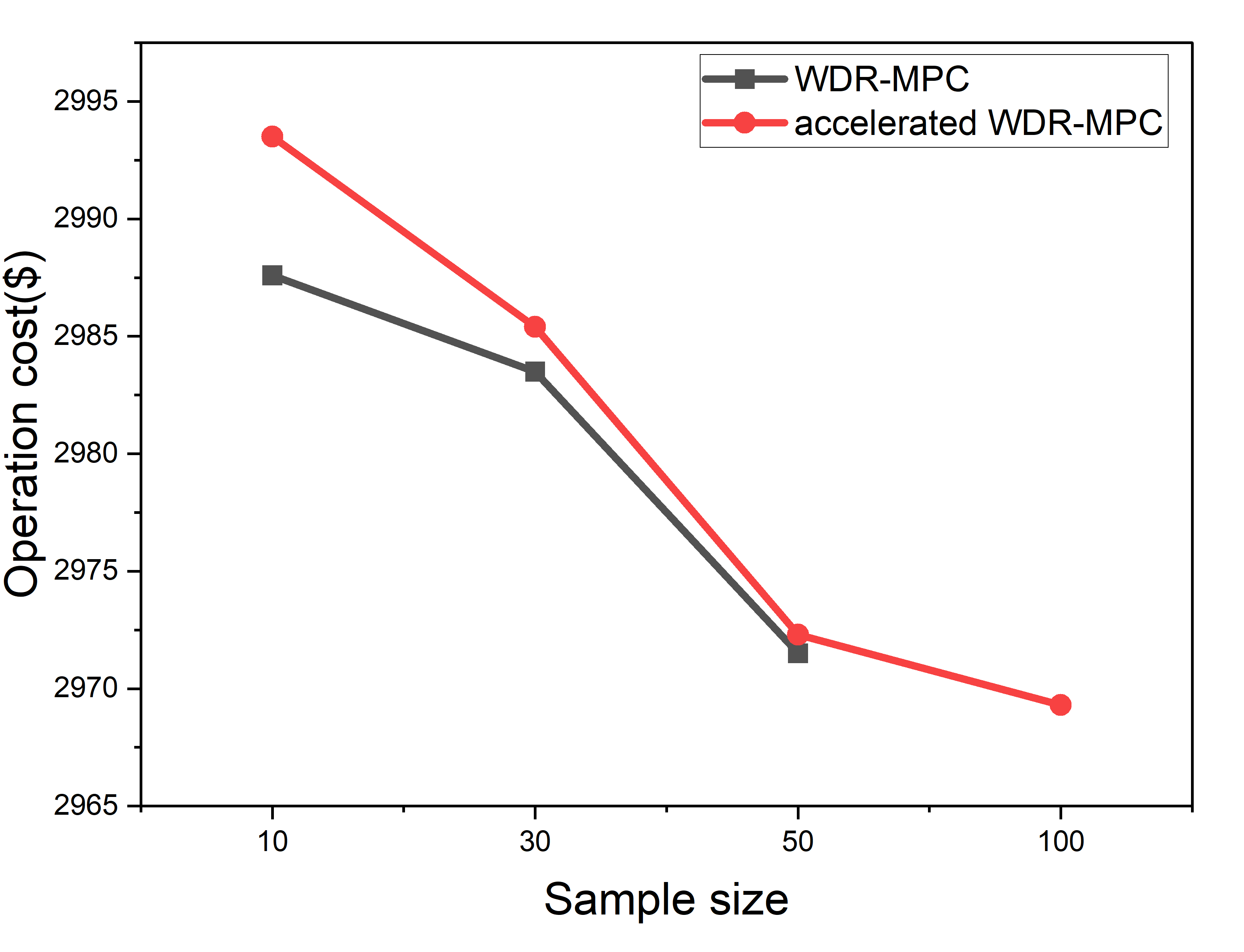}
    \caption{Comparison of WDR-MPC and accelerated WDR-MPC}
    \label{comparison of DRMPC and acce-DRMPC}
\end{figure}

Figure~\ref{comparison of DRMPC and acce-DRMPC} shows that the operational costs of both WDR-MPC and accelerated WDR-MPC tend to decrease as the sample size grows. Although an operational cost gap exists between WDR-MPC and accelerated WDR-MPC, the value is small and tends to diminish as the sample size increases. 

\begin{figure}[h]
    \centering
    \includegraphics[width = 0.9 \columnwidth]{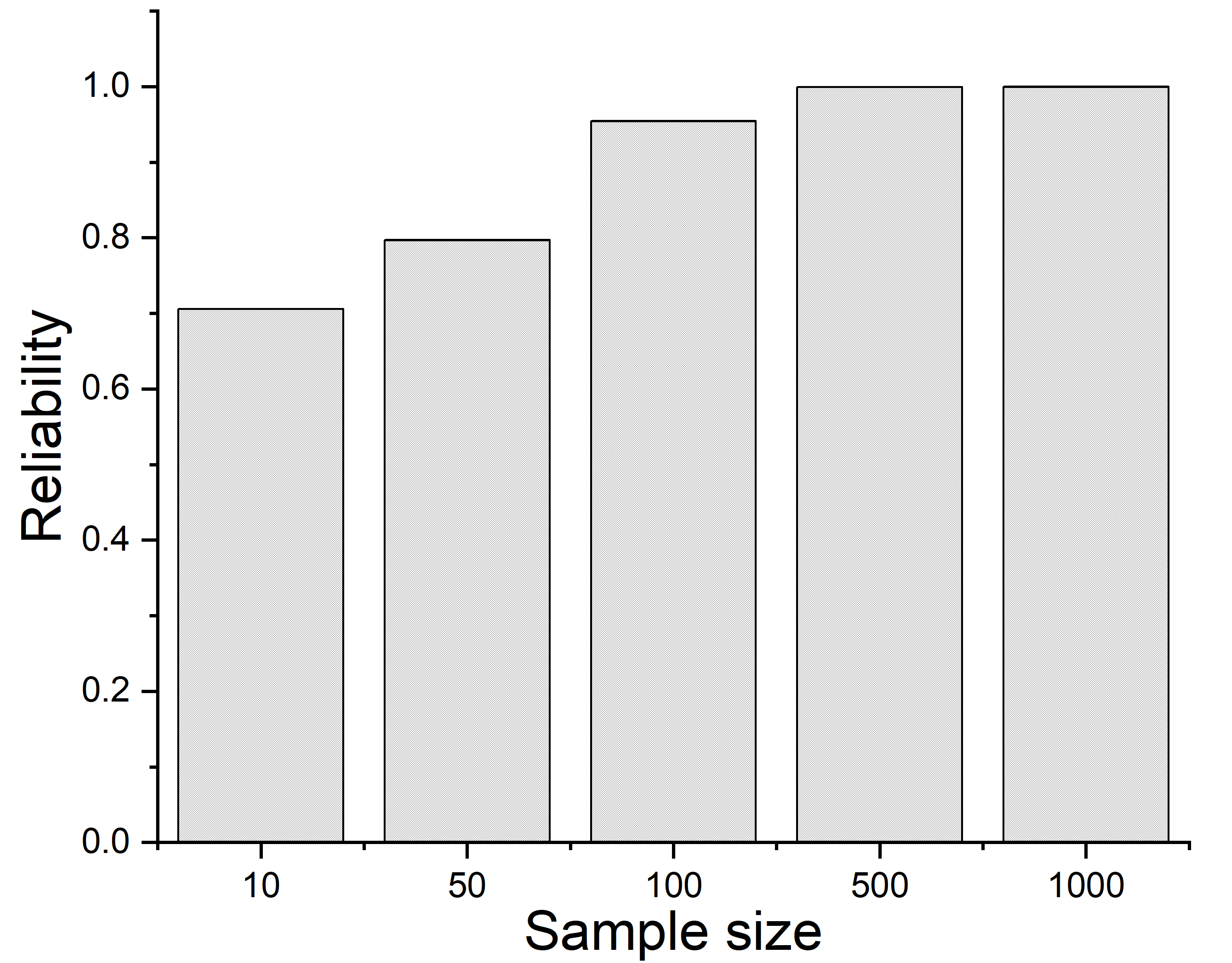}
    \caption{Reliability comparison of random disturbances in WDR-MPC}
    \label{Reliability}
\end{figure}

Figure \ref{Reliability} shows the results of the out-of-sample performance tests of WDR-MPC with random disturbances and dynamic uncertainty. We conducted these performance tests for 50 rounds with 5000 out-of-sample tests per round. The average reliability rate of 50 rounds is reported. The reliability rate represents the frequency of events where the optimal control actions still satisfy the constraints under another series of random disturbances $\left\{\omega(t)\right\}_{t=1}^{t=T_{l}}$. \textcolor{black}{The real RES power $\left\{\bm{\xi}(t)\right\}_{t=1}^{t=T_{l}}$ is randomly chosen from the testing set and the random disturbance series $\left\{\omega(t)\right\}_{t=1}^{t=T_{l}}$ are randomly drawn from an underlying uniform distribution over an interval with the center at the predicted charging demand of charging stations.} The radius $\epsilon$ and confidence levels $\beta$ were set to $0.001$ and $95\%$, respectively, and the sample size was set to $30$ with static uncertainty. The results show that, as the sample size grows larger, the reliability rate tends to increase to 100$\%$ while the violation rate decreases. 

\begin{figure}[h]
    \centering
    \includegraphics[width = 0.9 \columnwidth]{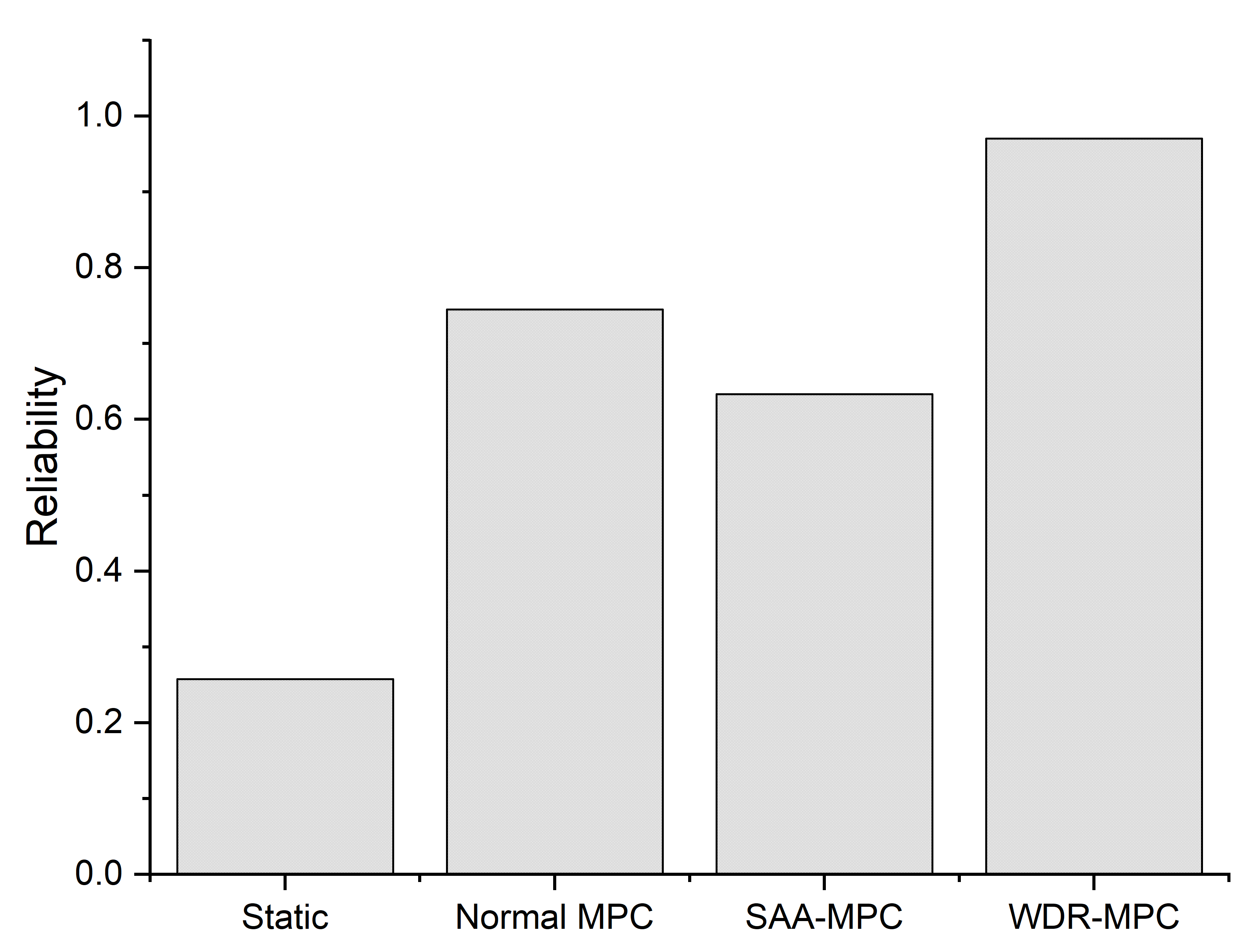}
    \caption{Comparisons between Static method, Normal MPC, SAA-MPC, and WDR-MPC for dynamic uncertainties}
    \label{Methods Comparison}
\end{figure}

\textcolor{black}{The efficiency of WDR optimization with static uncertainty has been well-documented; hence, we only compared the efficiency of WDR-MPC to other methods in a scenario of dynamic uncertainty. The choice of the prediction methods is not our main scope, and we simply used the sample average as the prediction method. The methods we compared are the Static method \cite{guo2018data1, guo2018data2}, Normal MPC \cite{li2022distributionally}, and SAA-MPC \cite{lee2022probabilistic}. The main difference between these methods lies in how they tackle dynamic uncertainty: 1) the static method treats random charging demand as flexible loads and solves traditional multi-stage WDR optimization for static uncertainties; 2) normal MPC uses the average samples as the prediction instead of constructing ambiguity tubes; 3) SAA-MPC constructs the ambiguity tubes and derives its bounds via sample average approximation and confidence level estimation; 4) our method (WDR-MPC) constructs ambiguity tubes via the WDR optimization method. The experiments involved out-of-sample performance tests with 500 samples from an unknown distribution. The radius $\epsilon$ was set to 0.001, and the confidence level $\beta$ was set to 95\%. The static uncertainty sample size was 30, and 500 for the dynamic uncertainty. As we can see in Figure \ref{Methods Comparison}, WDR-MPC gained an edge over the other three methods in the out-of-sample tests. This indicates the robustness of WDR-MPC in the scenario with dynamic uncertainty.}

\begin{figure}[h] 
    \centering
    \includegraphics[width= 0.9 \columnwidth]{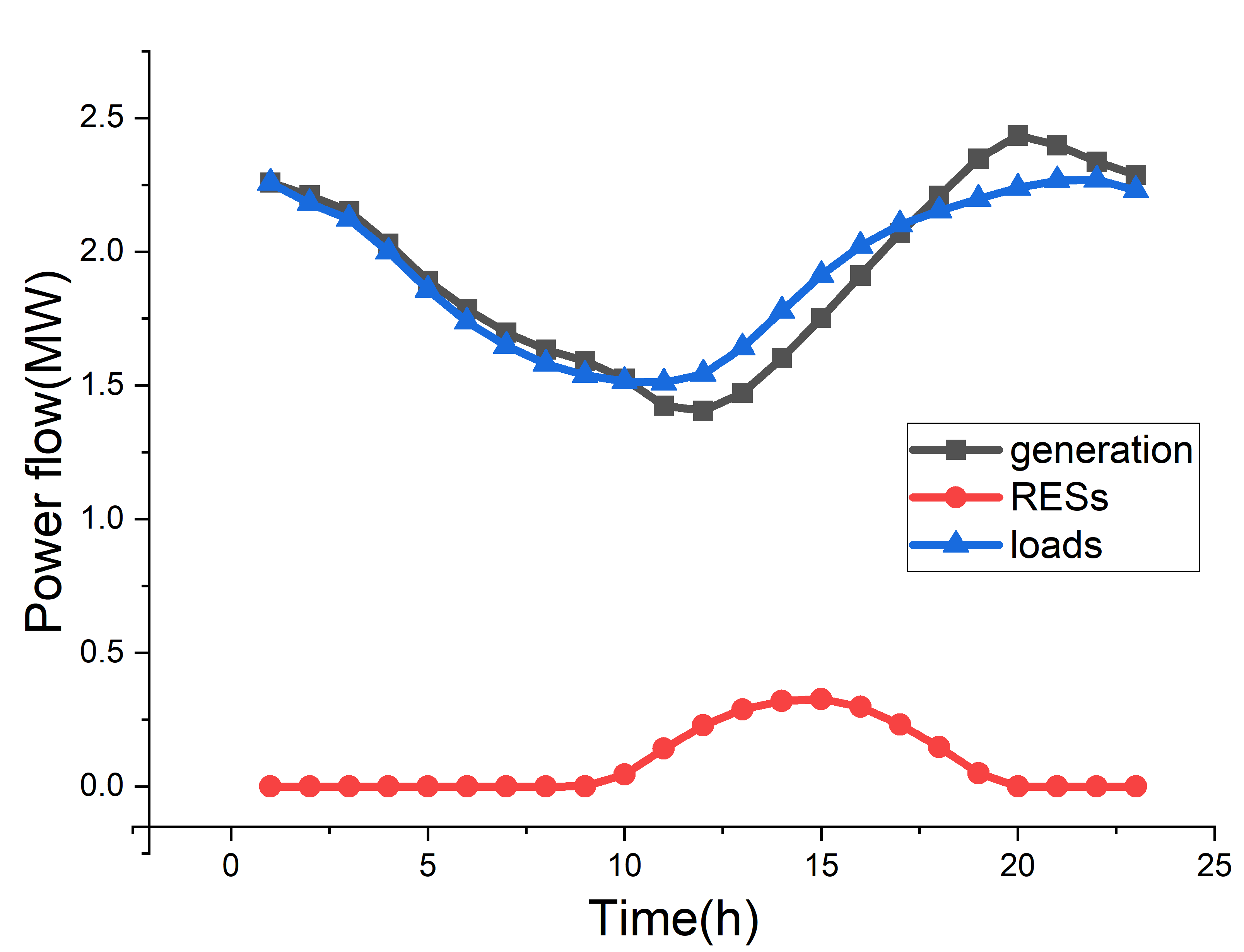}
    \caption{The power demands, power generation, and the RES generation of IEEE 38-bus system}
    \label{power fluctuation}
\end{figure}

The results of both the static and dynamic uncertainty scenarios were obtained with the parameter settings $\beta = 0.95,\epsilon = 0.001$. In our model, the voltage of each node varies within a small neighborhood but is far from the voltage limitations. Thus, we have omitted these parts of the results, providing just the figures for power flow. Figure \ref{power fluctuation} shows the power demands, power generation, and the RES power 
generation.
\subsection{IEEE 94-bus system}
The IEEE 94-bus system consists of 94 buses and 93 lines, with node 1 connected to a transmission system. Here, the grid model is modified by incorporating 7 distribution generators (Node 2, 6, 10, 12, 19, 20, and 26), 5 RES generators (Nodes 6, 10, 15, 19, and 25), as well as 2 charging stations (Nodes 2 and 6). 

In our simulation, we let $\underline{P}^{G}_{k}$, $\overline{P}^{G}_{k}$, $\underline{V}_{k}$ and $\overline{V}_{k}$ keep the same with the IEEE 38-bus system. The sample size for static uncertainty is chosen to be 30 and 500 for dynamic uncertainty. The Wasserstein radius for both stages is 0.001. We set the MPC time range $T_c = 8$, and the MPC time receding horizon is 16h. The confidence level $\beta$ is set to be 0.95. The computation time of accelerated WDR-MPC is 1057.2s for each MPC period, and WDR-MPC failed in this case due to a heavy computation burden arising from a more complex network.

\begin{figure}[h]
    \centering
    \includegraphics[width=0.9 \columnwidth]{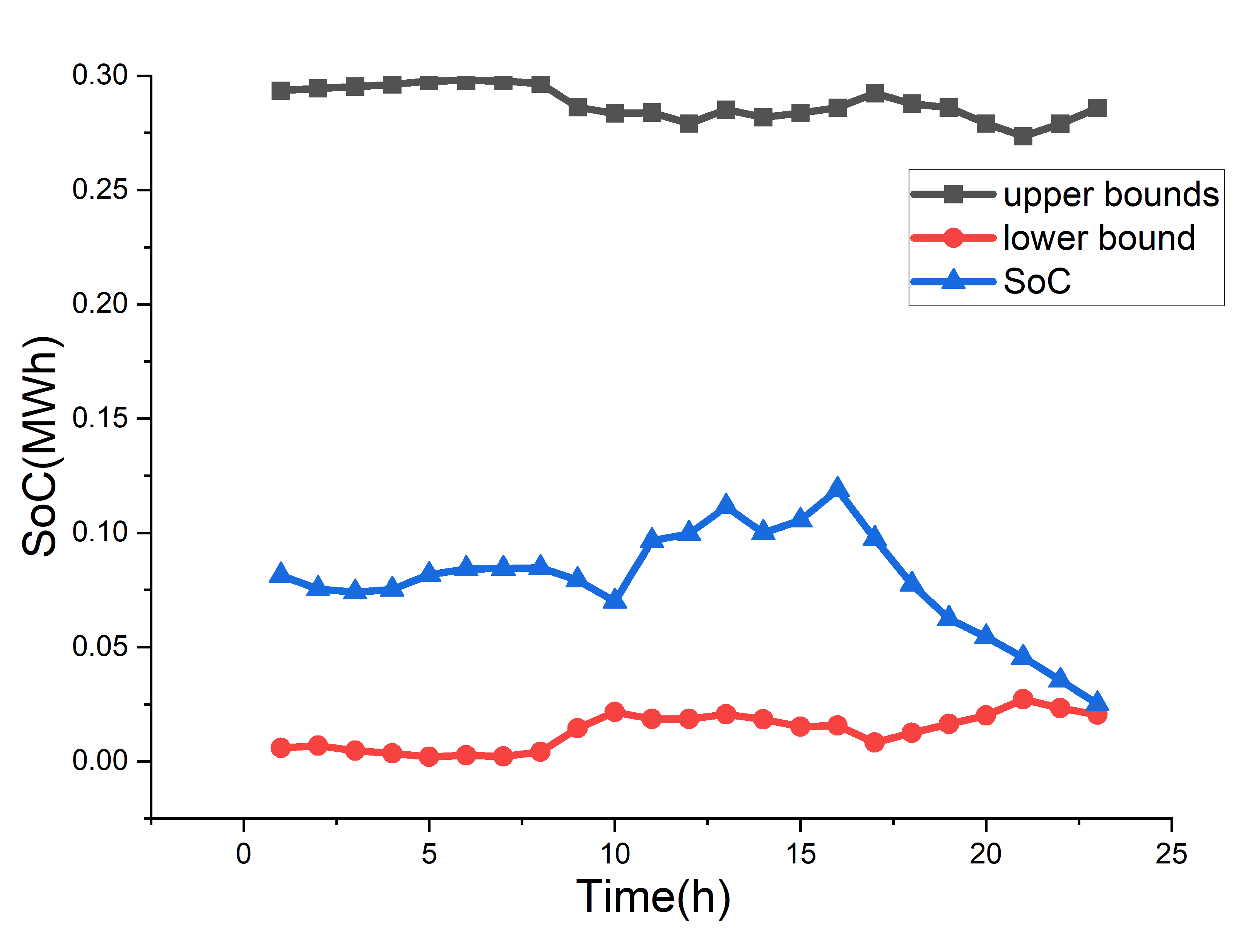}
    \caption{The fluctuation of SoC of IEEE 94-bus system}
    \label{SoC fluctuation94}
\end{figure}

\begin{figure}[h]
    \centering
    \includegraphics[width=1 \columnwidth]{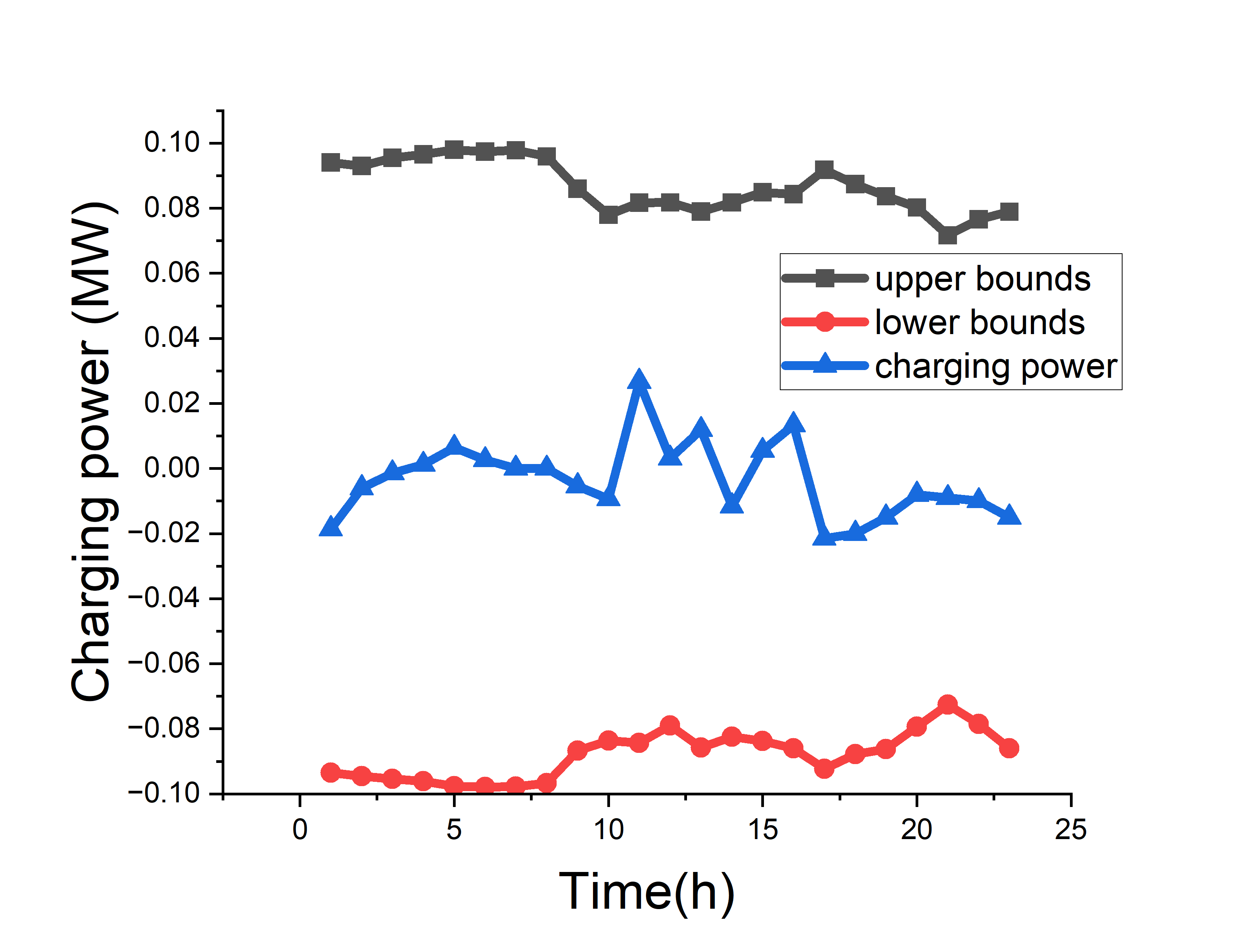}
    \caption{The fluctuation of charging power of IEEE 94-bus system}
    \label{charging power fluctuation94}
\end{figure}

Figure \ref{SoC fluctuation94} and Figure \ref{charging power fluctuation94} provide the fluctuations for the SoC, the charging power, and their corresponding upper/lower bounds in the simulations of one charging station. As we can see, the SoC and charging power are far from the upper/lower bounds at nearly all times. The reliability rate was almost 100\%, which further validates the efficiency of our method in a complex network. 
\vspace{-0.5cm}

\textcolor{black}{We also conducted numerical experiments to show comparison results for different degrees of risk ($1-\beta$). In these numerical experiments, the simulation results of different confidence levels $\beta$ are summarized in Table \ref{different beta reliability cmoparison}.}
\begin{table}[htbp!]
    \centering
    \caption{Simulation of different confidence level $\beta$}         
    \renewcommand{\arraystretch}{1.5}
    \begin{tabular}{|c|c|c|c|c|c|c|}
    \hline
    $\beta$ & 0.98 & 0.95 & 0.90\\
     \hline
     Reliability & 0.991 & 0.991 & 0.982\\
     \hline
     System Cost & 478.8 & 477.7 & 476.3 \\
     \hline
     \end{tabular}
\label{different beta reliability cmoparison}
\end{table}

\textcolor{black}{As $\beta$ becomes smaller, the degree of risk grows and therefore the reliability decreases. The system costs (not including risk costs) increase with the confidence level, because the system operator is willing to sacrifice some financial benefits to make less risky decisions.}

\textcolor{black}{Moreover, we conducted experiments under the settings with different numbers of charging stations to validate the scalability of our method. We ran four simulations with different numbers of charging stations: 1 charging station (Node 2), 2 charging stations (Nodes 2 and 6), 4 charging stations (Nodes 2, 6, 10 and 12), 6 charging stations (Nodes 2, 6, 10, 12, 19 and 20). The simulation results of different numbers of charging stations are summarized in the following Table \ref{different number of charging stations comparison}.
\begin{table}[htbp!]
    \centering
    \caption{Simulation of different number of charging stations}         
    \renewcommand{\arraystretch}{1.5}
    \begin{tabular}{|c|c|c|c|c|}
    \hline
    Charging station number & 1 & 2 & 4 & 6 \\
     \hline
     Offline time (s) & 496.82 & 1054.71 & 2011.11 & 3044.00\\
     \hline
     Online time (s) & 313.77 & 314.37 & 315.31 & 315.89\\
     \hline
     Reliability & 0.9990 & 0.9992 & 0.9905 & 0.9947 \\
     \hline
    \end{tabular}
\label{different number of charging stations comparison}
\end{table}
}

\textcolor{black}{As we can see in Table \ref{different number of charging stations comparison}, adding more charging stations will increase the offline optimization time, because we derive the ambiguity tube of each charging station in the offline optimization stage. We need to solve more separate but similar optimization problems as the number of charging stations grows. The increase in offline computation time can be alleviated by employing some parallel computing techniques. Table \ref{different number of charging stations comparison} further illustrates that the online computation time exhibits negligible variation. This is because more charging stations only add a few optimizing variables and constraints compared to existing optimizing variables. }


\textcolor{black}{The out-of-sample performance comparison of different numbers of charging stations is summarized in Table \ref{different number of charging stations comparison}. 
We set $\beta = 0.95$ and the sample size of dynamic uncertainty is 500. Other details are the same as in the previous experiments. As we can see, the out-of-sample performance does not vary much as the number of charging stations increases. The time consumption and reliability test comparison validate our method's reliability and scalability for multiple charging stations.}

\section{Conclusion}
In this paper, we have introduced WDR-MPC, a unified Wasserstein-based distributionally-robust MPC framework that is the first attempt to effectively handle both static uncertainty and dynamic uncertainty in smart grids. Through a well-designed two-stage algorithm, we successfully addressed these types of uncertainty. Our approach involves constructing ambiguity tubes and then deriving distributionally-robust bounds using the WDR optimization method. Moreover, we also devised a scalable acceleration method to alleviate the computational burden associated with WDR optimization. Numerical experiments validate the ability of the method to handle multiple types and instances of uncertainty. These experiments, which were conducted on real distribution networks, demonstrate the efficiency and effectiveness of our method. Our future work will focus on extending WDR-MPC to distributed scenarios with smart 
grids.

\section{APPENDIX}
\small
\subsection{Proof of Proposition 1}
    Let $f(\bm{y},\bm{\xi}) = \la \bm{a}(\bm{y}),\bm{\xi}\ra + \bm{b}(\bm{y})$ and let
    \begin{align}\nonumber
        F_{\beta}(\bm{y},\bm{\xi}) = \bm{\omega}+ \frac{1}{1-\beta}\int_{y\in\mathbb{R}^n} \left[ f(\bm{y},\bm{\xi})-\bm{\omega}\right]_{+}p(\bm{y})dy. 
    \end{align}
    Here we only give detailed derivation of $\bm{J}^{G}_{risk}$, the rest for $\bm{J}^{V}_{risk}$ is similar. 
    Based on \textbf{Theorem 1} in Rockafellar et al. \cite{rockafellar2000optimization}, we can obtain the loss function
    \begin{align} \nonumber
        \bm{J}_{risk,k}^G &= \bm{\omega}_1^G + \bm{\omega}_2^G + \frac{1}{1-\beta} \mathbb{E}^{\mathbb{P}} \left\{[\la \bm{\alpha}_{k}e,\bm{\xi}\ra + \bm{d}_{k}^G-\bm{\omega}_1^G]_{+} \right. \\
                        & + \left.[\la -\bm{\alpha}_{k}e,\bm{\xi}\ra -\bm{u}_{k}^G-\bm{\omega}_2^G]_{+} \right\} \nonumber\\
                        & = \mathbb{E}^{\mathbb{P}} \left\{\max_{j\in\{1,2,3,4\}}\left[ \la \bm{a}^G_{jk},\bm{\xi}\ra + \bm{b}_{jk}^G\right]\right\}, \nonumber
    \end{align}
    where $\bm{a}_{1k} = 0$, $\bm{a}_{2k} = \frac{1}{1-\beta}\bm{\alpha}_ke$, $\bm{a}_{3k} = \frac{-1}{1-\beta}\bm{\alpha}_ke$, $\bm{a}_{4k} = 0$, $\bm{b}_{1k} = \bm{\omega}_1^G + \bm{\omega}_2^G + \frac{1}{1-\beta}\left(\bm{d}_{k}^G-\bm{\omega}_1^G-\bm{u}_{k}^G-\bm{\omega}_2^G\right)$, $\bm{b}_{2k} = \bm{\omega}_1^G + \bm{\omega}_2^G + \frac{1}{1-\beta}\left(\bm{d}_{k}^G-\bm{\omega}_1^G\right)$, $\bm{b}_{3k} = \bm{\omega}_1^G + \bm{\omega}_2^G + \frac{1}{1-\beta}\left(-\bm{u}_{k}^G-\bm{\omega}_2^G\right)$ and $\bm{b}_{4k} = \bm{\omega}_1^G + \bm{\omega}_2^G$.

\subsection{Proof of Proposition 2}
    Consider the original WDR optimization problem \eqref{DRO}, where the Wasserstein metric is defined with $\ell_1$-norm; then we can express it in the following form:
    \begin{subequations}
    \begin{align}
    \min_{\bm{x} \in \mathcal{X}} \quad & \quad \sup_{\mathcal{P} \in \mathcal{B}_{\epsilon}} \mathbb{E}^{\mathcal{P}}\left(l(\bm{x},\bm{\xi})\right),\\
    \mathrm{s.t.} \quad & \quad \int_{\Xi^{2}}\Vert \bm{\xi}-\bm{\xi}'\Vert\Pi(d\bm{\xi},d\bm{\xi}') \leq \epsilon.
    \end{align}
    With the assumption that $\bm{\xi} \in \Xi = [\underline{\xi},\overline{\xi}]$, we can derive the dual of the above optimization problem:
    \begin{align}
    \min_{\bm{\lambda}, \bm{s}_i,\bm{x}} \quad &  \bm{\lambda}\epsilon+\frac{1}{N} \sum_{i=1}^{N} \bm{s}_i,\\
    \mathrm{s.t.} \quad &\sup_{\bm{\xi}}\left(l(\bm{x},\bm{\xi})-\bm{\lambda}\Vert \bm{\xi}-\hat{\xi}_{i}\Vert\right) \leq \bm{s}_i \label{s_i}\\
    & \bm{\lambda} \geq 0, \qquad i \in \{1,...,N_s\},\\
    & \bm{x} \in \mathcal{X},
    \end{align}
    \end{subequations}
    where $N_s$ is the sample size and $\Hat{\xi}_i$ denotes the historical sample point. Since $l(\bm{x},\bm{\xi})$ is convex with respect to $\bm{x}$ and $\bm{\xi}$, it is trivial to find that the supremum of $l(\bm{x},\bm{\xi})-\lambda\Vert \bm{\xi}-\hat{\xi}_{i}\Vert_1$ is achieved at the vertexes because $l(\bm{x},\bm{\xi})-\lambda\Vert \bm{\xi}-\hat{\xi}_{i}\Vert_1$ is convex on the interval $[\underline{\xi},\hat{\xi}_i]$ and $[\hat{\xi}_i,\overline{\xi}]$. Thus, we can eliminate the supremum and replace it with values at the three vertices, as follows: 
    \begin{subequations}
    \begin{align} \label{1-norm_2}
    \min_{\bm{\lambda},\bm{s}_i,\bm{x}} \qquad & \bm{\lambda}\epsilon+\frac{1}{N} \sum_{i=1}^{N} \bm{s}_i, \\
    \mathrm{s.t.} \qquad & l(\bm{x},\underline{\xi})+\bm{\lambda}(\underline{\xi}-\hat{\xi}_i)\leq \bm{s}_i, \\ 
    &  l(\bm{x},\overline{\xi})-\bm{\lambda}(\overline{\xi}-\hat{\xi}_i)\leq \bm{s}_i \\
    &  l(\bm{x},\hat{\xi}_i)) \leq \bm{s}_i, \\
    & \bm{\lambda} \geq 0, \bm{x} \in \mathcal{X}.
    \end{align}
    \end{subequations}
    Let $\eta(\bm{x},\bm{\xi}) = \dfrac{dl(\bm{x},\bm{\xi})}{d\bm{\xi}}$ and $\lambda_0 = \max\left\{\eta(\bm{x},\overline{\xi}),-\eta(\bm{x},\underline{\xi})\right\}$.
    When $\bm{\lambda} \geq \lambda_0$, since $l$ is convex for all $\bm{\xi} \in \left[\underline{\xi},\overline{\xi}\right]$, we have
    \begin{subequations}
    \begin{align}
        l(\bm{x},\bm{\xi}) & \geq l(\bm{x},\underline{\xi})+\bm{\lambda}(\underline{\xi}-\bm{\xi}),\\
        l(\bm{x},\bm{\xi}) & \geq l(\bm{x},\overline{\xi})-\bm{\lambda}(\overline{\xi}-\bm{\xi}).
    \end{align}
    \end{subequations}
    Thus we have 
    \begin{subequations}
    \begin{align}
    \min_{\bm{\lambda} \geq 0,\bm{x}\in \mathcal{X}} \qquad &  \bm{\lambda}\epsilon+\frac{1}{N}\sum_{i=1}^{N} l(\bm{x},\hat{\xi}_i),\\
    \mathrm{s.t.} \qquad\;\; & \eta(x,\overline{\xi}) \leq \bm{\lambda}, \\
    & -\eta(x,\underline{\xi}) \leq \bm{\lambda}.
    \end{align}
    \end{subequations}

\bibliographystyle{ieeetr}
\bibliography{final_WDR-MPC.bib}

\begin{IEEEbiography}[{\includegraphics[width=1in,height=1.25in,clip,keepaspectratio]{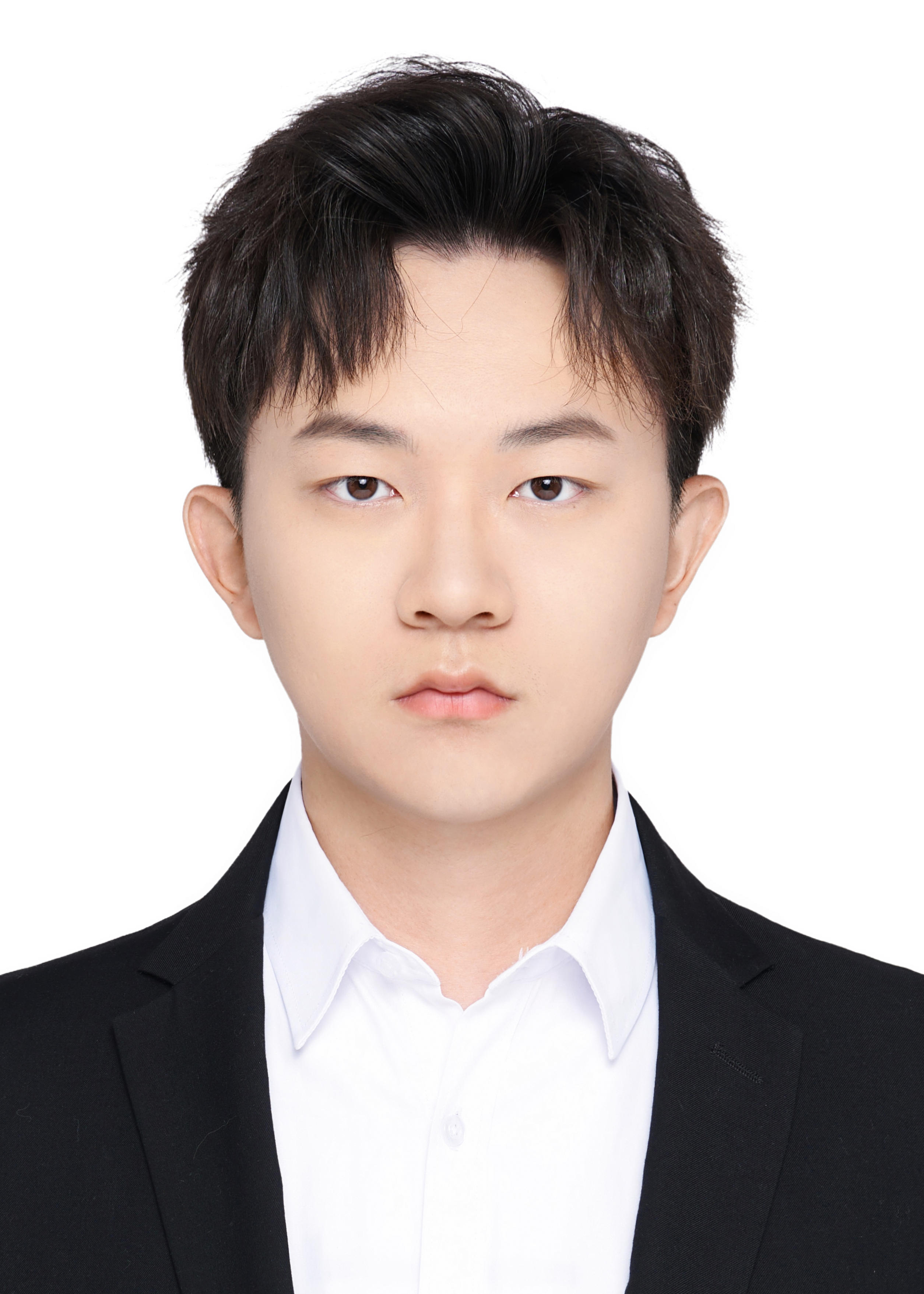}}]
{Qi Li} received the B.S. degree from the Institute of Mathematical Sciences, ShanghaiTech University, China, 2023. He is working towards the M.S.E degree in applied mathematics and statistics at Johns Hopkins University, USA. His current research interests include distributionally robust optimization and model predictive control. 
\end{IEEEbiography}

\begin{IEEEbiography}[{\includegraphics[width=1in,height=1.25in,clip,keepaspectratio]{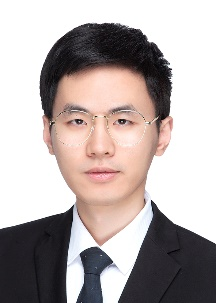}}]
{Ye Shi} (M’19) received Ph.D. degree from the school of Electrical and Data Engineering, University of Technology Sydney (UTS), Australia, 2018. He was a Research Assistant at the University of New South Wales, Australia from 2017 to 2019, and a Postdoctoral Fellow at UTS from 2019 to 2020. Since January 2021, Dr Shi has been an Assistant Professor in the School of Information Science and Technology at ShanghaiTech University. His research interests mainly focus optimization algorithms for Artificial Intelligence, Machine Learning and Smart Grid. Dr Shi was a recipient of the Best Paper Award at the 6th IEEE International Conference on Control Systems, Computing and Engineering in 2016. He serves as a Reviewer for many top-tier journals, such as IEEE TNNLS, TFS, JSAC, TSG, TPS, TII, TIE, etc; and many top-tier AI conference, such as ICML, NeurIPS, ICLR, CVPR, AAAI, etc. 
\end{IEEEbiography}

\begin{IEEEbiography}[{\includegraphics[width=1in,height=1.25in,clip,keepaspectratio]{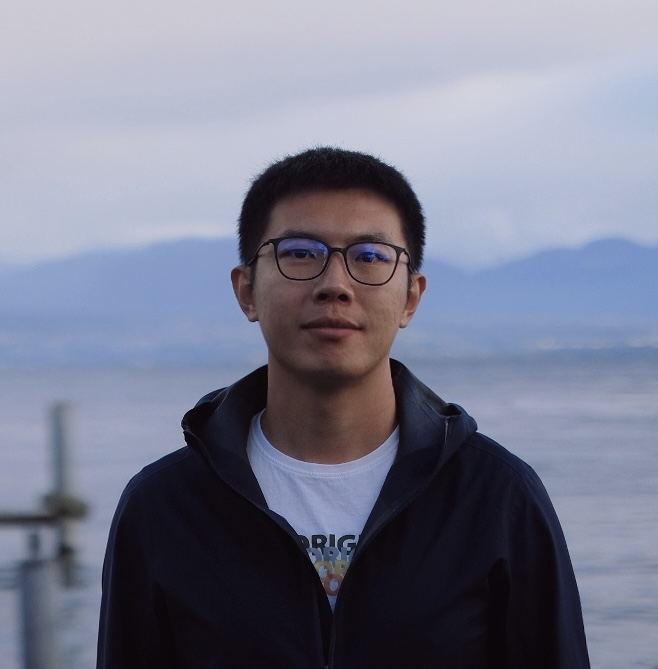}}]
{Yuning Jiang} (M'20) received the B.Sc. degree in electronic engineering from Shandong University, Jinan, China, in 2014, and the Ph.D. degree in information engineering from ShanghaiTech University, Shanghai, China, and the University of Chinese Academy of Sciences, Beijing, China, in 2020. He was a Visiting Scholar with the University of California at Berkeley (UC Berkeley), Berkeley, CA, USA, the University of Freiburg, Freiburg im Breisgau, Germany, and Technische Universität Ilmenau (TU Ilmenau), Ilmenau, Germany, during his Ph.D. study. He is currently a Postdoctoral Researcher with the Automatic Control Laboratory, École Polytechnique Fédérale de Lausanne (EPFL), Lausanne, Switzerland. His research focuses on learning- and optimization-based policy for operating complex systems, such as nonlinear autonomous systems (e.g., autonomous vehicles, robotics, and smart buildings), and large-scale multiagent systems (e.g., power and energy systems, IoT, and traffic networks).
\end{IEEEbiography}

\begin{IEEEbiography}[{\includegraphics[width=1in,height=1.25in,clip,keepaspectratio]{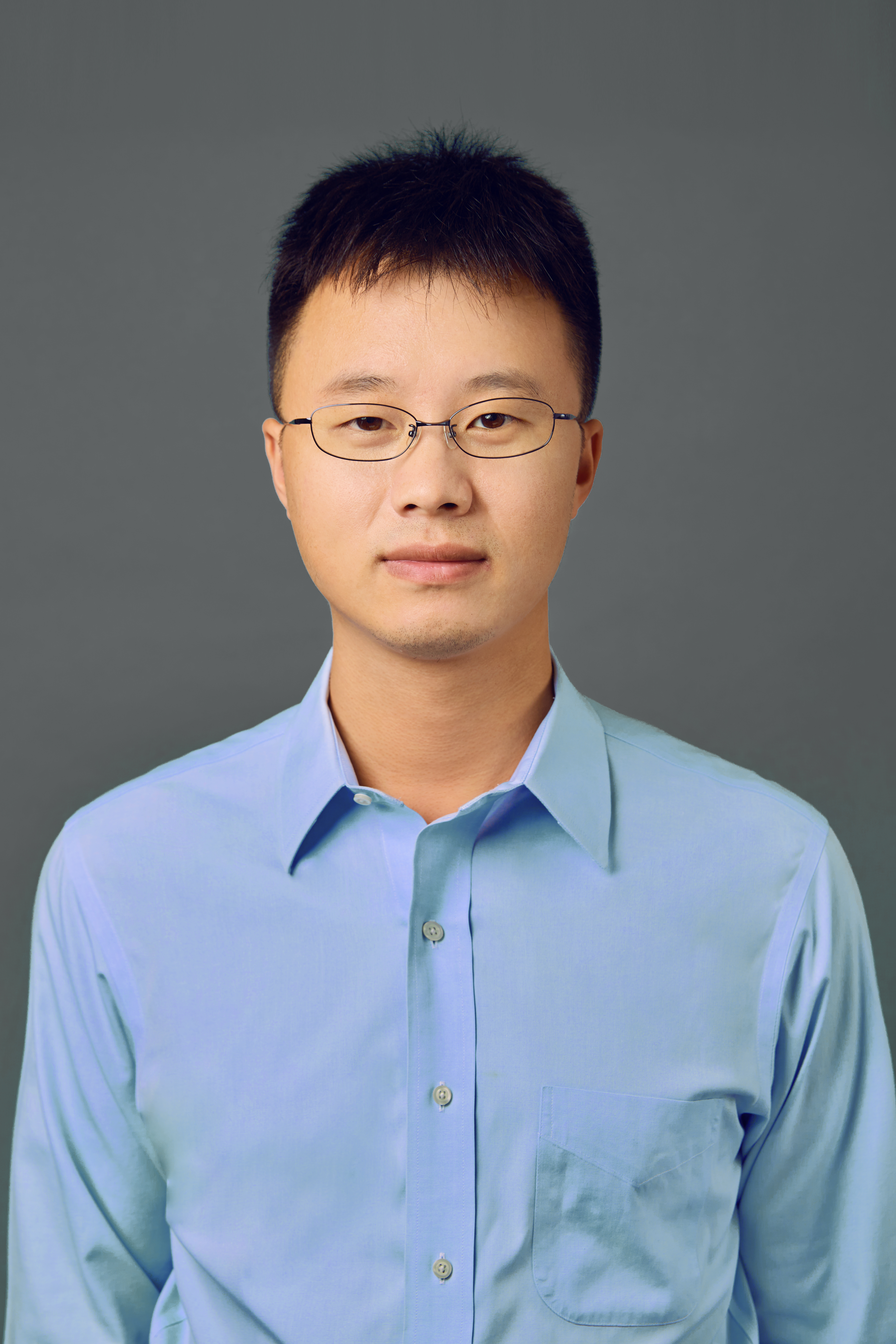}}]
{Yuanming Shi} (S’13-M’15-SM’20) received the B.S. degree in electronic engineering from Tsinghua University, Beijing, China, in 2011, and the Ph.D. degree in electronic and computer engineering from The Hong Kong University of Science and Technology, Hong Kong, in 2015.
Since September 2015, he has been with the School of Information Science and Technology, ShanghaiTech University, Shanghai, China, where he is currently a Full Professor. He visited the University of California at Berkeley,
Berkeley, CA, USA, from October 2016 to February 2017. His research areas include edge AI, federated edge learning, task-oriented communications, and satellite networks.
Dr. Shi was a recipient of the IEEE Marconi Prize Paper Award in Wireless Communications in 2016, the Young Author Best Paper Award by the IEEE Signal Processing Society in 2016, the IEEE ComSoc Asia–Pacific Outstanding Young Researcher Award in 2021, and the Chinese Institute of Electronics First Prize in Natural Science in 2022. He is also an editor of IEEE TRANSACTIONS ON WIRELESS COMMUNICATIONS, IEEE JOURNAL ON SELECTED AREAS IN COMMUNICATIONS, and JOURNAL OF COMMUNICATIONS AND INFORMATION NETWORKS. He is an IET Fellow.
\end{IEEEbiography}

\begin{IEEEbiography}[{\includegraphics[width=1in,height=1.25in,clip,keepaspectratio]{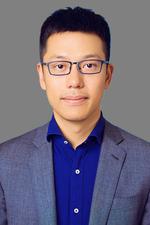}}]
{Haoyu Wang} ((S'12-M'14-SM'19) received the bachelor’s degree (Hons.) in electrical engineering from Zhejiang University, Hangzhou, China, in 2009, and the Ph.D. degree in electrical engineering from the University of Maryland, College Park, MD, USA, in 2014. In September 2014, he joined the School of Information Science and Technology, ShanghaiTech University, Shanghai, China, where he is currently an Associate Professor with tenure. In 2023, he visited the University of Cambridge, Cambridge, U.K., as a Visiting Academic Fellow. His research interests include power electronics, pulsed power supply, plug-in electric vehicles, and renewable energy systems.,Dr. Wang is an Associate Editor of IEEE Transactions on Industrial Electronics, IEEE Transactions on Transportation Electrification, and CPSS Transactions on Power Electronics and Applications. He is also a Guest Editor of IEEE Journal of Emerging and Selected Topics in Power Electronics
\end{IEEEbiography}

\begin{IEEEbiography}[{\includegraphics[width=1in,height=1.25in,clip,keepaspectratio]{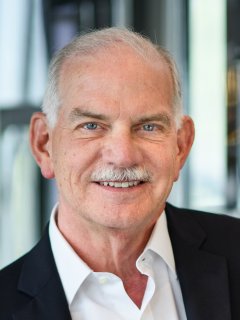}}]
{H. Vincent Poor} (S’72, M’77, SM’82, F’87) received the Ph.D. degree in EECS from Princeton University in 1977. From 1977 until 1990, he was on the faculty of the University of Illinois at Urbana-Champaign. Since 1990 he has been on the faculty at Princeton, where he is currently the Michael Henry Strater University Professor. During 2006 to 2016, he served as the dean of Princeton’s School of Engineering and Applied Science. He has also held visiting appointments at several other universities, including most recently at Berkeley and Cambridge. His research interests are in the areas of information theory, machine learning and network science, and their applications in wireless networks, energy systems and related fields. Among his publications in these areas is the book Advanced Data Analytics for Power Systems. (Cambridge University Press, 2021). Dr. Poor is a member of the National Academy of Engineering and the National Academy of Sciences and is a foreign member of the Royal Society and other national and international academies. He received the IEEE Alexander Graham Bell Medal in 2017
\end{IEEEbiography}
\end{document}